\pdfoutput=1
\documentclass[aps,twocolumn,superscriptaddress, prl]{revtex4-2}
\usepackage{graphicx}
\usepackage{dcolumn}
\usepackage{bm}
\usepackage{amsmath}
\usepackage{amssymb}
\usepackage{braket}
\usepackage[caption=false]{subfig}
\usepackage[colorlinks=true]{hyperref}
\usepackage{microtype}
\usepackage{comment}
\usepackage{xcolor}

\usepackage{physics}
\usepackage{mathrsfs}
\usepackage{comment}

\def\*#1{\mathbf{#1}}
\def\mc#1{\mathcal{#1}}
\def\mb#1{\mathbb{#1}}

\def\t#1{\text{#1}}

\def\mf#1{\mathfrak{#1}}

\def\bs#1{\boldsymbol{#1}}
\def\tt#1{\textit{#1.}|}

\def\grad{\bs\nabla}

\usepackage{ulem}

\begin{document}
\title{Topological hydrodynamics in spin-triplet superconductors}

\author{Chau Dao}
\thanks{These authors contributed equally to this work.}
\affiliation{Department of Physics and Astronomy and Bhaumik Institute for Theoretical Physics, University of California, Los Angeles, California 90095, USA}

\author{Eric Kleinherbers}
\thanks{These authors contributed equally to this work.}
\affiliation{Department of Physics and Astronomy and Bhaumik Institute for Theoretical Physics, University of California, Los Angeles, California 90095, USA}

\author{Bjørnulf Brekke}
\affiliation{Center for Quantum Spintronics, Department of Physics, NTNU - Norwegian University of Science and Technology, NO-7491 Trondheim, Norway}

\author{Yaroslav Tserkovnyak}
\affiliation{Department of Physics and Astronomy and Bhaumik Institute for Theoretical Physics, University of California, Los Angeles, California 90095, USA}


\date{\today}

\begin{abstract} 
    Due to the structure of the underlying SO(3) $\*d$-vector order parameter, spin triplet superconductors exhibit a bulk-edge correspondence linking the circulation of supercurrent to the bulk magnetic skyrmion density, giving rise to \textit{topological hydrodynamics} of magnetic skyrmions. To probe the interplay of charge and spin dynamics, we propose a blueprint for a spin-triplet superconducting quantum interference device (SQUID), which functions without a Josephson weak link. The triplet SQUID undergoes \textit{nonsingular} $4\pi$ phase slips, in which current relaxation is facilitated by spin dynamics that trace out a magnetic skyrmion texture. Inductively coupling the device to a tank circuit and probing the nonlinear supercurrent response via Oersted field measurements could provide an experimental signature of ferromagnetic spin-triplet superconductivity. 
\end{abstract}

\maketitle

\tt{Introduction}The potential for spin-triplet superconductors to uncover novel physical phenomena and further the development of quantum information technologies has spurred research efforts to experimentally identify new unconventional superconductors and validate the spin-triplet pairing~\cite{satoRPP17, ranSCI19,kasteningPRL06}. Ferromagnetic superconductors are a particularly interesting class of triplet superconductors that are not only robust to magnetic fields that would typically be deleterious to spin-singlet superconductivity, but also exhibit ferromagnetic order~\cite{aokiJPSJ19, machidaPRL01, samokhinPRB02,huxleyPRB01}. Candidate materials include uranium heavy-fermion compounds~\cite{aokiJPSJ19} and, more recently, two-dimensional moiré materials~\cite{liuNAT20,sharpeSCI19,kokkelerPRB23}. In these systems, spin and charge degrees of freedom are intertwined, as dictated by the structure of the underlying $\*d$-vector order parameter~\cite{leggettAOP74, supmat_dao25}. Changes in the supercurrent flow will generically be accompanied by magnetic dynamics and, vice versa, manipulation of the spin texture can give rise to charge dynamics~\cite{cornfeldPRR21,kimPRB23,poniatowskiPRL22, blountPRL79, brataasPRL04}.

\begin{figure}[t]
\includegraphics[width=0.8\linewidth]{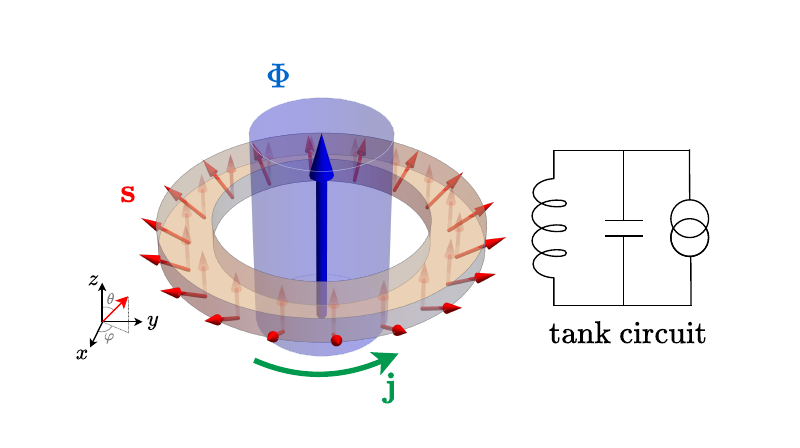}
\caption{Superconducting ring (left) with the spin orientation field $\*s$ (red arrows), circulating supercurrent $\*j$ (green arrow), and the magnetic flux (blue arrow), coupled to a tank circuit.}
\label{fig:1}
\end{figure}

\begin{figure}
    \centering
    \includegraphics[width=0.75
    \linewidth]{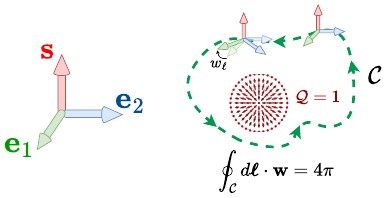}
    \caption{Sketch of the order parameter $\* d$ via the orthonormal triad $\{\*e_1,\*e_2,\*s\}$ (left) and the bulk-boundary correspondence between superconducting winding density $\*w$ and skyrmion charge $\mc Q$ (right).}
    \label{fig:2}
\end{figure}

In this Letter, we are motivated to explore the connection between charge and spin texture dynamics, with an emphasis on how magnetic topological charge transport can control superconducting current flow. Our focus is on a superconducting state characterized by a spatially-dependent SO(3) order parameter $\*d(\vb{r})$, see Fig.~\ref{fig:2}, which is momentum independent. This can potentially be realized in bilayer graphene as an $s$-wave, valley-singlet, spin-triplet superconductor~\cite{cornfeldPRR21}. Importantly, spin-triplet superconductors exhibit \textit{topological hydrodynamics} of magnetic skyrmions. Different from skyrmions in conventional magnets, spin-triplet skyrmions acquire a nonlocal character due to a bulk-edge correspondence linking the circulation of supercurrent to the bulk magnetic skyrmion density.

To illustrate the consequences of this physics, we propose a minimal setup for a spin-triplet superconducting quantum interference device (SQUID), shown in Fig.~\ref{fig:1}. Whereas spin-singlet SQUIDs require a Josephson weak link to facilitate \textit{singular} phase slips (where the order parameter vanishes), the spin-triplet SQUID can undergo \textit{nonsingular} phase slips in which the order parameter does not vanish~\cite{halperinIJMPB10}. This is possible because, as dictated by topology, a magnetic skyrmion flux induces a transverse change in the superconducting phase winding~\cite{merminPRL76,kimPRB23,cornfeldPRR21}. Readout can be performed by inductively coupling the spin-triplet SQUID to a tank circuit and probing the supercurrent via Oersted field measurements~\cite{tinkham95}. This can serve as a signature experiment for ferromagnetic spin-triplet superconductivity. Finally, we discuss a more complete dynamical theory that includes dissipation in the triplet superconductor via Gilbert damping of the spin degrees of freedom and Joule heating of the normal current.

\tt{Order parameter geometry}The link between charge and magnetic dynamics is rooted in the geometry of the $\*d$-vector order parameter. In the minimal model of a fully-spin polarized triplet superconductor, the order parameter can be expressed as $\*d(\*r,t) = d(\*e_1 + i\*e_2)/\sqrt{2}$, which is a complex vector field of modulus $d\equiv \sqrt{\*d\cdot\*d^*}$. 
The orientational degrees of freedom of $\*d$ are captured by
\begin{equation}
    \*s = \*e_1\times\*e_2,~~~\*w = (\grad \*e_2)\cdot\*e_1.
\end{equation}
We infer that $\*s$ is the spin orientation field since it satisfies the spin-commutation algebra (see below). Together with $\*{e}_1$ and $\*{e}_2$, it forms the spatially varying orthonormal triad $\{\*e_1,\*e_2,\*s\}$, see Fig.~\ref{fig:2}. In addition, 
$\*w(\*r,t)$ is the superconducting winding density~\footnote{Allowing for local gauge transformations, $\*w$ can be understood as a connection. In this Letter, we fixed the gauge, but we provide a gauge-independent formulation in the Supplemental Material~\cite{supmat_dao25}}, which tracks rotations of the $\*d$ vector about $\*s$~\cite{daoPRB25,kamienRMP02}. Assuming that the system is symmetric under global rotations of $\*e_1$ and $\*e_2$ about $\*s$ (corresponding to a mapping $\*d\rightarrow e^{i\phi}\*d$), $\*s$ and $\*w$ are the leading terms in $\*d$ satisfying this symmetry. Thus, they are the building blocks for our phenomenology.

\tt{Bulk-edge correspondence}The superconducting winding $\*w$ and the spin-orientation field $\*s$ are related through Stokes' theorem. For a 2-dimensional superconducting patch $\mc M$, we find
\begin{equation}\label{eq:bulkedge}
     \int_{\partial\mc M} \t d\bs\ell\cdot\*w =  \int_\mc M \t d^2r\,\varrho_\t{sk}(\*s),
\end{equation}
which has been written using the skyrmion density $\varrho_\t{sk}(\*s)\equiv \*s\cdot(\partial_x\*s\times\partial_y\*s)$. Equation (\ref{eq:bulkedge}) establishes a bulk-edge correspondence between the integrated winding along a boundary {$\partial\mathcal M$} and the number of skyrmions ${\cal Q}=\int_\mc M \t dr^2\, \varrho_\t{sk}/4\pi$, see Fig.~\ref{fig:2}.
As a skyrmion traverses the boundary $\partial \mc M$, there is a commensurate change in the superconducting winding $\*w$. Hence, the Stokes' theorem implies a topological continuity equation for magnetic skyrmions, $\partial_t\varrho_\t{sk} = -\grad\cdot\*j_\t{sk}$, where $\*j_\t{sk} = \*s\cdot[\partial_t\*s\times(\*z\times\grad)\*s]$ is the skyrmion flux. Note, the skyrmion continuity equation is also fulfilled in conventional Heisenberg magnets, which have an $S^2$ order parameter space. In that context, the continuity equation stems from the bulk homotopy $\pi_2(S^2) = \mb Z$, rather than from a Stokes' theorem~\cite{zouPRB19}. Therefore, those skyrmions are local objects that may fluctuate in and out of existence. In contrast, skyrmions in spin-triplet superconductors are inherently nonlocal objects that cannot be removed by local surgery and, thus, exhibit long-range interactions~\cite{knigavkoPRB99}.

Generally, by partially passing a skyrmion across an arbitrary 1-dimensional loop $\mc C$ (which may not necessarily be a boundary), the superconducting winding can be tuned to arbitrary values, a mechanism employed in Ref.~\cite{kimPRB23} \footnote{Conversely, the integrated superconducting winding along a closed loop is fixed if the spin texture is static. In this case, the one-dimensional winding density is a conserved quantity adhering to the topological continuity equation $\partial_tw_\ell = -\partial_\ell j_w$, where $j_w = -\*e_1\cdot\partial_t\*e_2$ is the winding flux. This continuity equation is spoiled by both singular and nonsingular phase slips.}. Fully moving a skyrmion across the loop engenders the nonsingular $4\pi$ phase slip that hallmarks spin-triplet superconductivity~\cite{cornfeldPRR21}. Formally, these nonsingular phase slips are smooth deformations of the $\*d$-vector field which preserve the topological sector of $\mc C$, classified according to the edge homotopy $\pi_1[\t{SO(3)}]= \mb Z_2$. When the spin texture is uniform along the loop, we recover integer-valued winding of the superconducting phase, $\oint_{\mc C} \t d\bs{\ell}\cdot\*w/2\pi = 2n + n_\t{top}$, where $n\in\mb Z$ and $n_\t{top} \in \{0,1\}$. The latter distinguishes the two homotopy classes of SO(3)~\cite{nakahara18}, corresponding to even or odd multiples of $2\pi$ winding~\footnote{If $\mc C$ can be identified with the boundary $\partial\mc M$ of a simply-connected superconducting patch $\mc M$, then $\mc C$ is restricted to the $n_\t{top}=0$ sector.}. 

\tt{Spin-triplet SQUID}The nonsingular phase slips rooted in the bulk-edge correspondence of  Eq.~(\ref{eq:bulkedge}) can be leveraged to devise the spin-triplet SQUID, which functions without a Josephson weak link. 
For this, we consider a one-dimensional superconducting ring of length $L$ threaded by external axial magnetic flux localized within the ring (shown in Fig.~\ref{fig:1}). The free energy $\mc F$ is constructed in terms of the spin orientation $\*s$ and winding density $w =\*e_1\cdot\partial_\ell\*e_2$ as
\begin{equation}\label{eq:freeenergysimple}
    \mc F[\*s,w] = \int\t d\ell\left\{\frac{A}{2}(\partial_\ell\*s)^2 + \frac{1}{2\chi_w}\left(w - \frac{q A_\ell}{\hbar c}\right)^2\right\},
\end{equation}
where $\ell$ is the arc length. The first term is the spin exchange energy with spin stiffness $A$, and the second term is the superconducting winding energy with $\chi_w$ being the winding compressibility and $q=-2e$ being the Cooper pair charge. 

Minimal coupling of the superconducting winding to the magnetic vector potential $A_\ell$ gives rise to the gauge-independent supercurrent $j = -c\delta_{A_\ell}\mc F=(q/\hbar\chi_w)(w{-}qA_\ell/\hbar c)$. Assuming that the current fulfills the incompressible continuity equation $\partial_\ell j=0$, integrating the supercurrent along $\ell$ and applying Stokes' theorem yields
\begin{align}\label{eq:current}
    j= -\frac{c}{L}\oint \t d{\ell} \, \delta_{A_\ell}\mc F=j_0 \left( 2\mc Q - \Phi/\Phi_0 + n_\t{top} \right),
\end{align}
where $j_0 \equiv 2\pi q/\hbar \chi_w L$ is the characteristic current. Here, the superconducting winding and the magnetic vector potential each give rise to a bulk curvature. The first is $\mc Q$, which is defined as the fictitious skyrmion charge enclosed by the SQUID and is dependent on the magnetic texture $\*s$. The second is the ratio of the magnetic flux $\Phi=\oint \t d\ell\,A_\ell$ to $\Phi_0 =  2\pi \hbar c/q$, the magnetic flux quantum~\cite{footnotesign,diracPRA31,tinkham95}. Finally, $n_\t{top}\in \{0,1\}$ is added to denote the topological sector of the 1-dimensional system. We aim to use the magnetic flux $\Phi$ as a handle to control the current on the boundary and drive skyrmion injection across the SQUID.

\tt{Quasistatic ansatz}For a sufficiently slowly varying magnetic flux, the response of the spin texture and supercurrent may be well-approximated by the quasistatic solution, which is given by the minimum of the free energy. Anticipating that the ground state free energy density is spatially homogeneous, we take an ansatz that the superconducting winding density $w$ is constant, and the spin $\* s$ uniformly sweeps out a cone around $\* z$ as we move along $\ell$, shown in Fig.~\ref{fig:1}. In spherical coordinates, the conical spin texture is described by a constant $z$ component $\cos \theta$ and in-$xy$-plane spin winding density $\partial_\ell \varphi=2\pi l/L$, where $l\in \mb Z$ is the integer-valued winding number that, by construction, can only change at $\theta=0,\pi$.

For this ansatz, the fictitious skyrmion charge becomes
\begin{align}\label{eq:Qs}
    \mc Q(\theta) \equiv n + l(1 - \cos\theta)/2,
\end{align}
where $n\in\mb Z$ is the integer number of fully enclosed fictitious skyrmions, and $l(1-\cos\theta)/2$ is the fraction that has not fully passed through the ring. By sweeping $\theta$ from 0 to $\pi$, we inject $l$ skyrmions across the 1-dimensional loop, which facilitates the nonsingular phase slip of $4\pi l$. 

Inserting the ansatz into the free energy of Eq.~(\ref{eq:freeenergysimple}), we obtain $ \mc F_{nl}(\Phi,\theta) = L_kj(\Phi,\theta)^2/2 + Jl^2\sin^2\theta/2$. The dependence on $\Phi$, $n$, and $l$ enters through $j$ from Eq.~(\ref{eq:current}) and $\mc Q$ from Eq.~(\ref{eq:Qs}). Here, $L_k=\chi_w L \hbar^2/q^2$ is the kinetic inductance and $J = 4\pi^2 A/L$ is the characteristic exchange energy. Similarly, $L_kj_0^2$ is the characteristic kinetic energy. For this discussion, we focus on the kinetic energy-dominated limit $\kappa =J/L_kj_0^2 < 1$, which is also the stability condition for our ansatz~\cite{supmat_dao25}.

In a more complete model, the magnetostatic energy must also be accounted for~\cite{supmat_dao25}. The predominant contribution to this energy arises from the Oersted field produced by the supercurrent. This can be included in $\mc F_{nl}$ by the simple replacement $L_k\rightarrow L_k + L_g$, where $L_g$ is the geometric inductance of the ring. Hence, the generated Oersted field favors the limit $\kappa<1$. A less significant contribution is the magnetic field from the dipoles, which can be small compared to the Oersted field if the circumference $L$ is much larger than the width of the ring~\cite{pengJMM04}.

\begin{figure*} 
    \centering
      \includegraphics[width=0.85\linewidth]{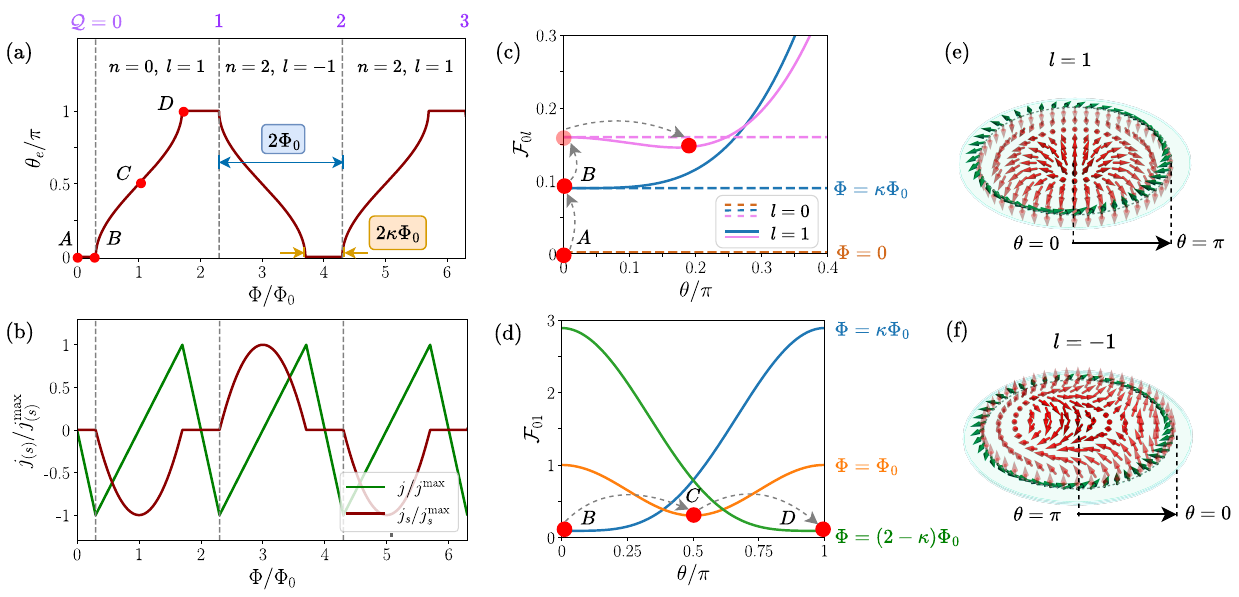}
    \caption{(a) Plot of the energetic minimum $\theta_e$ as magnetic flux $\Phi$ is varied. Gray dashed lines indicate different integer values of skyrmion charge $\mc Q_s$. (b) Plot of the normalized spin current $j_s/j_s^\t{max}$ (red) and electric current $j/j^\t{max}$ (green) as a function of $\Phi$. (c) Depiction of switching from $\mc F_{00}$ to $\mc F_{01}$ as the critical magnetic flux $\kappa \Phi_0$ is reached. Dashed lines represent $\mc F_{00}(\Phi,\theta)$ and solid lines represent $\mc F_{01}(\Phi,\theta)$. Opaque red dots mark stable points and the translucent red dot marks the unstable fixed point. (d) Evolution of the free energy $\mc F_{01}(\Phi,\theta)$. Red circles mark the energetic minima. (a)--(d) are plotted with $\kappa = 0.3$. (e) and (f) depict positive skyrmion charge swept out by the spin texture on the ring as $\theta_e$ traverses from 0 to $\pi$ and $\pi$ to 0. (e) shows $l=1$ spin winding and (f) shows $l=-1$ spin winding. Green arrows indicate $4\pi l$ superconducting phase winding. }
    \label{fig:3}
\end{figure*}

\tt{Nonlinear response}Within the quasistatic ansatz, the nonlinear response of the SQUID to the magnetic flux is determined by the steady-state spin texture $\theta_e$, which fulfills  $\partial_\theta\mc F_{nl}(\Phi,\theta_e) = 0$. The poles at $\theta_e = 0 $ and $\theta_e = \pi$ are always fixed points, which may be stable or unstable. Away from the poles, the extrema are 
\begin{equation}\label{eq:equilibrium}
     \theta_e = \cos^{-1}\left[\frac{2n+n_\t{top}+l-\Phi/\Phi_0}{(1-\kappa)l}\right],
\end{equation}
which, for valid values of $n$, $l$, and $\Phi$, are stable energetic minima. 
Once the equilibrium spin polar angle $\theta_e$  is determined for a given magnetic flux $\Phi$, the supercurrent follows directly from Eq.~\eqref{eq:current}.

As a concrete example, see Fig.~\ref{fig:3}, we initialize a system with $\Phi = 0$ ($A$) in the ``even" $n_\t{top} = 0$ topological sector so that the ground state has zero supercurrent, $j=0$, and spin texture is uniformly aligned along $\vb{s}_0=\vb{z}$ with $\theta_e=0$.
As the magnetic flux $\Phi$ is increased, negative supercurrent builds up as a diamagnetic response until it reaches the maximum value $-j^\t{max} =-\kappa j_0$ at the critical flux $\kappa \Phi_0$ ($B$). Here, the spin texture with $\theta_e=0$ becomes unstable. Upon further increasing $\Phi$, it becomes energetically favorable to relax the supercurrent by changing the spin winding from $l=0$ to $l=1$ and to initiate the nonsingular phase slip, see Fig.~\ref{fig:3}{\color{red}(c)}. The polar angle $\theta_e$ sweeps from the north pole $\theta_e=0$ ($B$) to the equator $\theta_e=\pi/2$ ($C$) to the south pole $\theta_e=\pi$ ($D$), see Fig.~\ref{fig:3}{\color{red}(d)}, whereupon the spin texture on the ring has traced out a full positive skyrmion charge at $(2-\kappa)\Phi_0$, see Fig.~\ref{fig:3}{\color{red}(e)}. This completes a nonsingular $4\pi$ phase slip of the superconducting phase, bringing the supercurrent to $+j^\t{max}$. Now, the diamagnetic response of the superconductor repeats 
until $-j^\t{max}$ is reached at the critical value $(2+\kappa)\Phi_0$. At this point, the sweeping reverses from $\theta_e=\pi$ to $\theta_e=0$ with a magnetic winding $l=-1$, which, again,  traces out one full positive skyrmion charge at $(4-\kappa)\Phi_0$, see Fig.~\ref{fig:3}{\color{red}(f)}, engendering a second nonsingular $4\pi$ phase slip. This pattern repeats with periodicity $4\Phi_0$. 
We remark that, without Zeeman or spin-orbit coupling, the free energy in Eq.~(\ref{eq:freeenergysimple}) is invariant under global rotation of $\*s$. Hence, the magnetic flux-induced switching can happen between arbitrary antipodal points $\*s_0$ and $-\*s_0$.
 
In the spin-triplet SQUID, the response of the supercurrent to the magnetic flux is nonlinear even without a Josephson junction, as seen in Fig.~\ref{fig:3}{\color{red}(b)}. While the electric current's sawtooth behavior~\cite{tinkham95} is reminiscent of the current response in the Little-Parks effect in $\t{U}(1)$ superconductors~\cite{littlePRL62}, the current of the triplet superconductor relaxes continuously via nonsingular phase slips rather than singular phase slips. This gives rise to distinct falling and rising edge slopes $-j^\t{max}/\Phi_0 \kappa$ and $j^\t{max}/\Phi_0(1-\kappa)$, respectively, which, along with the $2\Phi_0$ periodicity in the magnetic flux~\cite{cornfeldPRR21,tinkham95}, is an indicator of triplet superconductivity. Singular phase slip events, on the other hand, would relax the superconducting winding by $2\pi$, and are distinguishable from the nonsingular phase slips since they halve the flux periodicity to $\Phi_0$~\cite{tinkham95}. The supercurrent response could be measured by an inductively coupled tank circuit, operating analogously to the readout of the $\text{U}(1)$ rf SQUID~\cite{jackelJLP75,tinkham95}. 

Besides a charge supercurrent, the triplet superconductor also sustains a spin current $j_s=-A\vb{z} \cdot\left(\vb{s}\times \partial_\ell \vb{s}\right)$ polarized in the $z$ direction~\cite{soninAIP10,zhuARX25}. At equilibrium, the electric and spin currents are related by 
\begin{align}
\frac{j}{j_s}= \frac{q}{\hbar}\frac{\cos\theta_e}{\sin^2\theta_e},
\end{align}
which follows from Eqs.~\eqref{eq:current}-\eqref{eq:equilibrium} and holds for all values of $\Phi$ and $\theta\neq0,\pi$. We find that the spin current exhibits an oscillatory behavior with periodicity $4\Phi_0$, twice that of the supercurrent, see Fig.~\ref{fig:3}{\color{red}(b)}.  
Notably, the magnitude of the spin current is maximized when the charge current is zero. Vice versa, the magnitude of the charge current reaches its maximal value when the spin texture is uniform, meaning zero spin current.

\tt{Dissipation}For a more complete treatment, we include dissipation of the triplet superconductor due to, e.g., Gilbert damping and Joule heating of the normal current. The set of noncanonical degrees of freedom are chosen to be $\bold{X}= \{\rho,\*s,w\}$, where we have included the total charge density $\rho=q d^2$ to account for both superconducting and normal electrons~\footnote{From the microscopic theory, it is known that the Poisson bracket between $\rho$ and the $\*d$-vector is $\{\rho(\ell),\*d(\ell')\} = iq\*d(\ell)\delta(\ell-\ell')/\hbar$~\cite{supmat_dao25}. When we specify the canonical $\*d$-vector Poisson brackets, we also find that $q d^2$ shares the same Poisson structure (hence, the same equation of motion) as $\rho$. Thus, $qd^2$ is identified with the total charge density. Similarly, $s=\hbar d^2$ is the total spin density.}. Physically, this corresponds to the normal electrons being enslaved to the superconducting electrons. When constructing the equations of motion, we must make sure that the dissipative terms preserve the topological bulk-edge correspondence $\partial_t \mc Q=\oint \mathrm{d} \ell j_\text{sk}/4\pi$, where $j_\t{sk} \equiv  \*s\cdot(\partial_t\*s\times\partial_\ell\*s)$. Furthermore, they must maintain the spin length,  $\*s^2 = 1$,  and obey charge conservation. To this end, we propose a variant of Hamilton's equations~\cite{chaikin95,felderhofJCP11},
\begin{subequations} \label{eq:hamilton}
\begin{equation}
    \dot X_i(\ell,t) = \sum_j\int \mathrm{d}{\ell'}\big\{X_i(\ell,t),X_j(\ell',t)\big\}\mf{F}_j(\ell',t).
\end{equation}
Here, we have defined the generalized force 
\begin{equation}
    \bs{\mf{F}} \equiv \delta_{\*X}\mc F[\*X] + \delta_{\dot {\*X}}\mc R[\dot{\*X}]
\end{equation}
\end{subequations}
to be the sum of the nondissipative force stemming from the free energy $\mc F$ and the dissipative force, which is derived from the Rayleigh functional $\mc R$. Remarkably, a generic Rayleigh functional $\mc R[\dot{\*X}]$ constructed from velocities $\dot{\*X}$ complies with the Poisson structure of the system and maintains the three structural invariants detailed above. In contrast, constructing the dissipation functional $R[\delta_{{\*X}}\mc F]$ from thermodynamic forces $\delta_{\*{X}}\mc F$~\cite{felderhofJCP11} may violate all three constraints~\footnote{The construction of $\mc R$ in terms of velocities $\dot{X}_i$ or forces $\delta_{X_i} \mc F $ are only equivalent when the relation $\dot{X}_i(\ell,t)=\sum_j\int \mathrm{d}{\ell'}\big\{X_i(\ell,t),X_j(\ell',t)\big\}\delta_{X_j(\ell',t)} \mc F $ is invertible, as is the case in canonical systems. In general, the map is noninvertible if there exist Casimir invariants~\cite{morrisonRMP98} (e.g., $\vb{s}^2$). While the velocity construction guarantees conservation of Casimir invariants, the more general force construction may not~\cite{supmat_dao25}.}. 

Once the Poisson brackets, the free energy, and the Rayleigh functional are specified, the equations of motion naturally follow. The Poisson brackets follow from that of the order parameter $\vb{d}$, $\{d_i^*(\ell),d_j(\ell')\} = i\delta_{ij}\delta(\ell-\ell')/\hbar$ and $\{d_i(\ell),d_j(\ell')\} =0$~\cite{cornfeldPRR21}, which can be motivated from a microscopic theory~\cite{supmat_dao25, bulou21}. We find~\cite{supmat_dao25,cornfeldPRR21,volovik09}
\begin{subequations}\label{eq:pb}
    \begin{align}
    \{s_i(\ell),s_j(\ell')\}&=\epsilon_{ijk}s_k\delta(\ell-\ell')/s,\label{eq:spinalgebra}\\    
    \{w(\ell),\rho(\ell')\} &=-q\partial_\ell\delta(\ell - \ell')/\hbar,\label{eq:10c}\\
    \left\{w(\ell),\*s(\ell^\prime) \right\}&= \partial_\ell \*s  \,\delta(\ell-\ell^\prime)/s,\label{eq:6d}
\end{align}
\end{subequations}
and all other Poisson brackets of $\rho,w$, and $\*s$ evaluate to zero~\footnote{Mention that in 2D $\vb{w}$ with itself is interesting.}.
Equation \eqref{eq:spinalgebra} allows, in retrospect, the identification of $\vb{s}$ as the spin-orientation field and $s\equiv \hbar d^2$ as the saturated spin density. Equation \eqref{eq:10c} is the counterpart of the Poisson bracket between phase and charge in a $U(1)$ superconductor. Finally, Eq.~\eqref{eq:6d} captures the linkage between superconducting winding and spin texture inherent to the $\vb{d}$ vector order parameter.

The free energy is given by Eq.~\eqref{eq:freeenergysimple}, with the addition of the nonelectrostatic contribution $\int \t d\ell \, (\rho -\rho_0)^2/2\chi_c$, where $\chi_c$ is the charge compressibility. In the Weyl gauge, the electrostatic contribution is absent and the nonelectrostatic contribution captures small fluctuations of $\rho$ away from the equilibrium charge density $\rho_0$.   
Finally, the Rayleigh functional is constructed from the gauge-independent velocities $\partial_t\*s$ and $\partial_t(w - qA_\ell/\hbar c)$: 
\begin{align}\label{eq:rayleighan}
    \mc R = \int \t d\ell\left\{\frac{\alpha s}{2} \left(\partial_t\*s\right)^2 + \frac{\sigma}{2}\left[\partial_t\left(\frac{\hbar w}{q} - \frac{A_\ell}{c}\right)\right]^2\right\}.
\end{align}
The first term corresponds to Gilbert damping with damping coefficient $\alpha$ and the second term describes Joule heating with conductivity $\sigma$. 
Using Eqs.~(\ref{eq:hamilton})--(\ref{eq:rayleighan}), the equations of motion are
\begin{subequations}\label{eq:equationsofmotion}
    \begin{align}
        \partial_t\rho&=-\partial_\ell \mf{J},\label{eq:drhodt}\\
        \partial_t\*s  &= \*s\times \*H_\t{eff} - \alpha\*s\times\partial_t\*s - \mf{J}\partial_\ell\*s/\rho,\label{eq:dsdt}\\
        \partial_t w &= -{q \partial_\ell \mu}/{\hbar}+j_\text{sk},\label{eq:dvdt}
    \end{align}
\end{subequations}
and have been written using the constitutive relation 
\begin{align}\label{eq:chargecurrent}
  \mf{J} = j + \sigma\mc E,~~~\text{where}~~~\mc E =E_\ell - \partial_\ell\mu + \hbar j_\t{sk}/q.
\end{align}
Here, $\*H_\t{eff} = -\delta_{\*s}{\mc F}/s$ is the effective magnetic field and $\mu = \delta_\rho \mc F$ is the chemical potential. The equations of motion describe the charge continuity equation in Eq.~(\ref{eq:drhodt}), the Landau-Lifshitz-Gilbert equation in Eq.~(\ref{eq:dsdt}), and the rate of change of winding in Eq.~(\ref{eq:dvdt}) which corresponds to the spin-triplet version of the first London equation~\cite{kimPRB23}. From the charge continuity equation (\ref{eq:drhodt}), $\mf{J} = - c \delta_{A_\ell} {\cal F}- c \delta_{\dot{A}_\ell} {\cal R}$ is identified as the total charge current, which is driven by the electromotive force $\mc E$. The latter has contributions from the electric field $E_\ell=-\partial_tA_\ell/c$, the gradient of $\mu$, and the transverse skyrmion flux directed towards the center of the loop $j_\text{sk}$. 

In accordance with Onsager reciprocity, the total current $\mf{J}$ induces an adiabatic spin-transfer torque on the spin orientation $\*s$, see Eq.~(\ref{eq:dsdt}). Importantly, there is no term in either the free energy or the Rayleigh function that couples the spin texture and the superconducting winding, which, in conventional magnets, would be necessary to produce a spin-transfer torque. Rather, in triplet-superconductors, the spin-transfer torque (and motive force) appears due to the nontrivial Poisson bracket between $w$ and $\*s$. This captures the essence of the topological linkage between skyrmion and charge currents~\cite{onsagerPR31,landau80}.

A linearized analysis of Eqs.~\eqref{eq:equationsofmotion} suggests that our quasistatic ansatz used to study the spin-triplet SQUID is stable even with Gilbert damping and Joule heating, as long as we ensure the kinetic energy dominated regime, $\kappa<1$. For $\kappa>1$, the quasistatic ansatz during the nonsingular phase slip process can no longer be assumed. Nevertheless, on topological grounds, we still expect switching of magnetization to facilitate relaxation of the supercurrent~\cite{supmat_dao25}. 

\tt{Conclusion}This work opens a pathway to employ techniques developed for spintronics on a spin-triplet superconductor platform. Due to the suppression of Joule heating, spin-triplet superconducting spintronics could enable efficient charge current control of spin textures. Moreover, the bulk-edge correspondence between supercurrent and magnetic skyrmions provides a handle on topological charge dynamics that would be otherwise absent in conventional magnets. The nonlocality of magnetic skyrmions in spin-triplet superconductors potentially makes them suitable for energy storage or long-range signal transport applications.

\smallskip

\begin{acknowledgments}
\tt{Acknowledgments}This work is supported by NSF under Grant No. DMR-2049979. B.B. acknowledges support from the Research Council of Norway through the Centres of Excellence funding scheme, Project No. 262633, ”QuSpin." C.D. and E.K. thank Shane Kelly for many fruitful discussions.
\end{acknowledgments}

\bibliography{ms.bib}

\begin{thebibliography}{45}
\bibitem [{foo()}]{footnotesign}%
  \BibitemOpen
  \href@noop {} {}\bibinfo {note} {Note, for $q = -2e<0$, the magnetic flux quantum $\Phi_0$ as well as $j_0$ are negative.}\BibitemShut {Stop}%
\end{thebibliography}

\end{document}


\preprint{APS/123-QED}

\title{Supplemental Material for \\ ``Topological hydrodynamics in spin-triplet superconductors"}

\author{Chau Dao}
\affiliation{Department of Physics and Astronomy and Bhaumik Institute for Theoretical Physics, University of California, Los Angeles, California 90095, USA}

\author{Eric Kleinherbers}
\affiliation{Department of Physics and Astronomy and Bhaumik Institute for Theoretical Physics, University of California, Los Angeles, California 90095, USA}

\author{Bjørnulf Brekke}
\affiliation{Center for Quantum Spintronics, Department of Physics, NTNU - Norwegian University of Science and Technology, NO-7491 Trondheim, Norway}

\author{Yaroslav Tserkovnyak}
\affiliation{Department of Physics and Astronomy and Bhaumik Institute for Theoretical Physics, University of California, Los Angeles, California 90095, USA}

\maketitle

In this Supplemental Material, we provide (1) a discussion of the $\*d$-vector order parameter and the microscopic Poisson brackets, (2) a variant of Hamiltonian formalism to address dissipation in constrained systems, (3) a derivation of the equations of motion for $\rho,\*s,\*w,\*A,\*E$, (4) a discussion on topological hydrodynamics in spin-triplet superconductors, (5) a derivation of the linearized modes of the spin-triplet SQUID and a stability analysis of the quasistatic solution, and (6) a characterization of the spin-triplet SQUID response in the $\kappa>1$ regime, with and without easy-axis anisotropy.

\section{1. Poisson brackets}
\subsection{(i) $\*d$-vector order parameter}

In this Letter, we are interested in fully spin-polarized triplet superconductivity, which can occur in twisted bilayer graphene and twisted double bilayer graphene. In these systems, electrons  are  endowed with a valley degree of freedom $\tau\in\{\*K,\*K'\}$~\cite{schaibleyNRM2016,kleinherbersPRB23} 
in addition to the spin degree of freedom $\sigma\in\{\uparrow,\downarrow\}$. 

In general, the superconducting order parameter can exhibit spatial dependence as well as momentum dependence. This \textit{mixed representation} can be captured by performing a Wigner-Weyl transform on the Cooper pair field operator $\psi^{\tau}_{\sigma}(\*r)\psi^{\tau^\prime}_{\sigma^\prime}(\*r^\prime)$ which is composed of two electron field operators with spin $\sigma$ and $\sigma^\prime$,  valley $\tau$ and $\tau^\prime$, and position $\*r$ and $\*r'$, respectively. We find that the Wigner transform is given by
\begin{align}
    \Psi_{\sigma\sigma^\prime}^{\tau \tau^\prime}(\*r,\*p) = 
    \int \frac{\mathrm{d}\bs{\mf r}}{(2\pi)^{d}}\,e^{-i\*p\cdot\bs{\mf r}}\psi^{\tau}_{\sigma}(\*r + \bs{\mf r}/2)\psi^{\tau^\prime}_{\sigma^\prime}(\*r-\bs{\mf r}/2),
\end{align}
where $\*r$ is the ``center-of-mass" coordinate, $\bs{\mf r}$ is the relative position between the electrons, and $\hbar \*p$ is the momentum conjugate to $\bs{\mf r}$.
Note that due to the Pauli principle, this operator fulfills
\begin{align}
    \Psi_{\sigma\sigma^\prime}^{\tau \tau^\prime}(\*r,\*p)=-\Psi_{\sigma^\prime \sigma}^{\tau^\prime \tau}(\*r,-\*p),
\end{align}
that is, flipping the spin $\uparrow\,\leftrightarrow\, \downarrow$, the valley $\*K \leftrightarrow \*K'$, and the momentum $\hbar \*p\leftrightarrow-\hbar\*p$ generates an overall minus sign. 
To comply with the Pauli principle, we assume for the superconducting ground state the expectation value 
\begin{align}
   \ev{\Psi_{\sigma\sigma^\prime}^{\tau \tau^\prime}(\*r,\*p)} = \sqrt{\frac{n}{2}}\*d(\*r,\*p) \cdot \left(\bs{\sigma}\, i\sigma_y\right)_{\sigma\sigma^\prime}\left(i\tau_y\right)^{\tau\tau^\prime},
\end{align}
which corresponds to a spin-triplet, valley-singlet order parameter with $\*d(\*r,\*p)=\*d(\*r,-\*p)$. 
Hence, the valley singlet can even enable an s-wave character with no dependence on $\*p$. Furthermore, we have introduced the Cooper pair density $n=(2\pi)^{-d}\int_{\t{shell}} \t{d}\*p$ by integrating over the approriate shell around the Fermi surface~\footnote{For a conventional superconductor, the the shell around the Fermi energy has a characteristic width given by the Debye frequency of the phonons that mediate the attractive interaction of electrons.}. Inverting the relation, we obtain
\begin{align} \label{eq:suppdvecrp}
\*d(\*r,\*p) =\frac{\sqrt{2}}{4\sqrt{n}} \left(-i\sigma_y \bs{\sigma}\right)_{\sigma'\sigma}\left(-i\tau_y\right)^{\tau' \tau}\ev{\Psi_{\sigma\sigma^\prime}^{\tau \tau^\prime}(\*r,\*p)}. 
\end{align}
Specifically, this results in~\cite{bulou21,leggettRMP75,leggettAOP74,kimPRB23} \begin{subequations}\label{supeq:Deltatod}
\begin{align}
    d_x&= \frac{1}{\sqrt{2}}\ev{\Psi^{\*K\*K'}_{\downarrow\downarrow} - \Psi^{\*K\*K'}_{\uparrow\uparrow}},\\
    d_y &= \frac{-i}{\sqrt{2}} \ev{\Psi^{\*K\*K'}_{\downarrow\downarrow} + \Psi^{\*K\*K'}_{\uparrow\uparrow}},\\
    d_z &= \frac{1}{\sqrt{2}}\ev{\Psi^{\*K\*K'}_{\uparrow\downarrow} + \Psi^{\*K\*K'}_{\downarrow\uparrow}}.
\end{align}
\end{subequations}
Marginalizing out the momentum $\hbar \*p$, we can identify the order parameter $\*d(\*r)$ used in the Letter:
\begin{align}\label{eq:suppmarginalize}
  \*d(\*r)=\int\limits_\t{shell}  \t{d}\*p\, \*d(\*r,\*p).
\end{align} 
On the other hand, marginalizing out the center-of-mass coordinate $\*r$ yields the well-known superconducting gap function
\begin{align}
     \Delta_{\sigma\sigma'}(\*p)= V_0\int \mathrm{d}\*r  \ev{\Psi_{\sigma\sigma^\prime}^{{\*K}\*K'}(\*r,\*p)} 
    &=V_0\ev{ a_{\*p\sigma}b_{-\*p\sigma'}},
\end{align}
where we have introduced the Fourier-transformed electron operators 
\begin{subequations}
\begin{align}
    a_{\*p\sigma}&=\int\frac{ \mathrm{d}{\* r}}{(2\pi)^{d/2}} e^{-i\*p \cdot \* r} \psi^{\*K}_{\sigma}(\*r),\\
    b_{\*p\sigma}&=\int\frac{ \mathrm{d}{\* r}}{(2\pi)^{d/2}} e^{-i\*p \cdot \* r}\psi^{\*K'}_{\sigma}({\* r}),
\end{align}
\end{subequations}
for valley $\*K$ and $\*K'$, respectively. $V_0$ is the electron-electron interaction which we have assumed, for simplicity, to have no structure in momentum, spin, and valley.

\subsection{(ii) Microscopic Poisson brackets}
In this section, we want to microscopically motivate the classical Poisson brackets of the $\* d$ vector order parameter given by 
\begin{equation}\label{eq:supppoissondvec}
    \{d_i^*(\*r), d_j(\*r')\} = \frac{i}{\hbar}\delta_{ij}\delta(\*r-\*r'),~~~\{d_i(\*r),d_j(\*r')\}=0.
\end{equation}
Similar Poisson brackets are found in Refs. \cite{cornfeldPRR21,kimPRB23},  which were introduced on phenomenological grounds.

Our strategy is to evaluate the quantum commutation relations of the Wigner-transformed Cooper pair operator $\Psi_{\sigma\sigma^\prime}^{{\*K}\*K'}(\*r,\*p)\equiv\Psi_{\sigma\sigma^\prime}(\*r,\*p)$, where we drop the valley indices from now on.
Then, we employ Dirac's correspondence principle to produce the classical Poisson brackets. Finally, we integrate over the momenta $\*p$ and $\*p'$ and obtain the classical Poisson brackets of $\*d(\*r)$.

We start by Fourier transforming the field operator via
\begin{align}\label{supeq:wigner}
\begin{split}
    \Psi_{\sigma_1\sigma_2}(\*r,\*p)&= \int \frac{\t{d}\bs{\mf r}}{{(2\pi)^{d}}}\, e^{-i\bs{\mf r}\cdot\*p}\left(\int \frac{\t{d}\*k_1}{(2\pi)^{d/2}}\frac{\t{d}\*k_2}{(2\pi)^{d/2}}e^{i(\*k_1+\*k_2)\cdot\*r}e^{i(\*k_1-\*k_2)\cdot\bs{\mf r}/2}a_{\*k_1\sigma_1}b_{\*k_2\sigma_2}\right)\\
    &= \frac{1}{(2\pi)^{2d}}\int  \t{d}\*k_1 d\*k_2\,e^{i(\*k_1+\*k_2)\cdot\*r}a_{\*k_1\sigma_1}b_{\*k_2\sigma_2}\int \t{d}\bs{\mf r} \,e^{-i\bs{\mf r}\cdot[\*p - (\*k_1-\*k_2)/2]}\\
    &= \frac{1}{(2\pi)^{d}}\int  \t{d}\*k_1d\*k_2\,e^{i(\*k_1+\*k_2)\cdot\*r}\delta(\*p - \*q)a_{\*k_1\sigma_1}b_{\*k_2\sigma_2}.
\end{split}
\end{align}
In the final line, we have defined $\*q = (\*k_1- \*k_2)/2$ as the relative momentum between the two electrons. 
The electron field operators adhere to the standard fermionic anticommutation relations
\begin{align}
\begin{split}
    \{a_{\*k\sigma}, a_{\*k'\sigma'}\} &= \left\{b_{\*k\sigma},b_{\*k'\sigma'}\right\} = 0,\\
    \{a_{\*k\sigma}, b_{\*k'\sigma'}^\dagger\} &= \left\{a_{\*k\sigma},b_{\*k'\sigma'}\right\} = 0,\\
    \{a_{\*k\sigma},a^\dagger_{\*k'\sigma'}\} &= \{b_{\*k\sigma},b^\dagger_{\*k'\sigma'}\} = \delta(\*k - \*k')\delta_{\sigma\sigma'}.
\end{split}
\end{align}
The commutation of the Wigner transformed operators is expressed as
\begin{align}\label{supeq:commutationbig}
    [\Psi_{\sigma_1\sigma_2}(\*r,\*p),\Psi^\dagger_{\sigma_1'\sigma_2'}(\*r',\*p')] = \frac{1}{(2\pi)^{2d}}\int \t d\*k_1 \t d\*k_2 \t d\*k_1' \t d\*k_2'\,e^{i(\*k_1 + \*k_2)\cdot\*r}e^{-i(\*k_1' + \*k_2')\cdot\*r'}\delta(\*p-\*q)\delta(\*p'-\*q')\left[a_{\*k_1\sigma_1}b_{\*k_2\sigma_2},(a_{\*k_1'\sigma_1'}b_{\*k_2'\sigma_2'})^\dagger\right]
\end{align}
which can be simplified by using 
\begin{align}
\left[a_{\*k_1\sigma_1}b_{\*k_2\sigma_2},(a_{\*k_1'\sigma_1'}b_{\*k_2'\sigma_2'})^\dagger\right] = \delta(\*k_1-\*k_1')\delta(\*k_2-\*k_2')\delta_{\sigma_1\sigma_1'}\delta_{\sigma_2\sigma_2'} - a_{\*k_1'\sigma_1'}^\dagger
a_{\*k_1\sigma_1}\delta(\*k_2-\*k_2')\delta_{\sigma_2\sigma_2'} - b_{\*k_2'\sigma_2'}^\dagger b_{\*k_2\sigma_2}\delta(\*k_1-\*k_1')\delta_{\sigma_1\sigma_1'}.
\end{align}
We arrive at 
\begin{equation}\label{eq:commutatorpsi}
   \left[\Psi_{\sigma_1\sigma_2}(\*r,\*p),\Psi^\dagger_{\sigma_1'\sigma_2'}(\*r',\*p')\right]= {(2\pi)^{-d}}{\delta(\*r - \*r')\delta(\*p - \*p')\delta_{\sigma_1\sigma_1'}\delta_{\sigma_2\sigma_2'}} +D_{\sigma_1\sigma_2\sigma_1'\sigma_2'}(\*r,\*r',\*p,\*p'),
\end{equation}
where the first term captures the bosonic nature of the Cooper pairs, while the second term is the deviation operator $D_{\sigma_1\sigma_2\sigma_1'\sigma_2'}$ stemming from the  Pauli exclusion principle of the underlying electrons~\cite{combescot15}. This term is neglected from here on. Similarly, we can derive 
\begin{equation}\label{eq:commutatorpsi}
   \left[\Psi_{\sigma_1\sigma_2}(\*r,\*p),\Psi_{\sigma_1'\sigma_2'}(\*r',\*p')\right]= 0,
\end{equation}
which is exact without corrections. Now, we can use the classical to quantum correspondence ${[\cdot, \cdot]}\rightarrow {i\hbar}\{ \cdot, \cdot \}$, and we obtain
\begin{subequations}
\begin{align}\label{eq:commutatorpsi}
   \{\langle{\Psi^{\phantom{\dagger}}_{\sigma_1\sigma_2}(\*r,\*p)\rangle},\langle{\Psi^\dagger_{\sigma_1'\sigma_2'}(\*r',\*p')}\rangle\}&={-i(2\pi)^{-d}} {\delta(\*r - \*r')\delta(\*p - \*p')\delta_{\sigma_1\sigma_1'}\delta_{\sigma_2\sigma_2'}}/\hbar \\  \{\langle{\Psi_{\sigma_1\sigma_2}(\*r,\*p)\rangle},\langle{\Psi_{\sigma_1'\sigma_2'}(\*r',\*p')\rangle}\} &= 0. 
\end{align}
\end{subequations}
As a next step, we use Eq.~\eqref{eq:suppdvecrp}, and arrive at
\begin{subequations}
\begin{align}\label{eq:commutatordrp}
   \left\{d_i(\*r,\*p),d^*_j(\*r',\*p')\right\}&=\frac{-i\hbar}{(2\pi)^{d} n} \delta_{ij}{\delta(\*r - \*r')\delta(\*p - \*p')} \\  \left\{d_i(\*r,\*p),d_j(\*r',\*p')\right\}&=0,
\end{align}
\end{subequations}
where we used the relation $\tr(\sigma_y \sigma_i \sigma_j \sigma_y)=2\delta_{ij}$. After integrating out the momenta $\*p$ and $\*p'$ and using Eq.~\eqref{eq:suppmarginalize}, 
we obtain the desired Poisson brackets
\begin{subequations}
\begin{align}\label{eq:commutatordr}
   \left\{d^*_i(\*r),d_j(\*r')\right\}&={i} \delta_{ij}{\delta(\*r - \*r')}/\hbar, \\  \left\{d_i(\*r),d_j(\*r')\right\}&=0,
\end{align}
\end{subequations}
where we used $n=(2\pi)^{-d}\int_{\t{shell}} \t{d}\*p$.

\subsection{(ii) Derived Poisson brackets}
The $\*{d}$ order parameter is a complex vector that we parameterize by
\begin{align}
    \*{d}(\*r) = \frac{d(\*r)}{\sqrt{2}} e^{i\phi(\*r)}\left[{\*{e}}_1(\*r) + i\,{\*{e}}_2(\*r)\right],\label{parametrization}
\end{align}
where the real vectors ${\*{e}}_1$ and ${\*{e}}_2$ are normalized, $\abs{\*e}_1=\abs{\*e}_2=1$. This leaves us with six degrees of freedom: the phase $\phi$, the magnitude $d$, and two orientational degrees of freedom each for $\*e_1$ and $\*e_2$.
The spin orientation field can be derived from the $\* d$ vector via~\cite{cornfeldPRR21}
\begin{align}
    \*{s}=\frac{i\*{d}\times \*{d}^*}{d^2}={\*{e}}_1 \times {\*{e}}_2.
\end{align}
Below, we show that $\*s$ indeed follows the usual spin algebra. 
Hence, for a fully spin-polarized superconductor with $\abs{\*s}=1$, we infer that $\*{e}_1\cdot {\*{e}}_2=0$. In this case, the $\* d$ vector also fulfills the gauge redundancy $e^{i\chi} R_{\* s}(\chi) \*d=\*d$, that is, a combination of a local phase transformation with a local rotation around $\* s$ leaves the order parameter unchanged. For a fully spin-polarized triplet superconductor, we are now left with four degrees of freedom described by the modulus squared $d^2$ and the triad $\{\*e_1,\*e_2,\*s\}$. Alternatively, we could choose the modulus squared $d^2$, the superconducting phase $\phi$, and the spin orientation $\* s$.  

Next, we define the winding density
\begin{align}\label{eq:winding}
    w_i=\frac{1}{2id^2}\left({\*{d}^* \cdot \partial_i \*{d} - \*{d} \cdot \partial_i \*{d}^*}\right)
   &= \partial_i \phi + {\*{e}_1}\cdot \partial_i {\*{e}_2},
\end{align}
where ${\*{e}_1}\cdot \partial_i {\*{e}_2}$ takes the form of a connection. In the Letter, we choose the gauge $\phi=0$. 
Assuming the order parameter does not vanish anywhere, $d^2\neq0$, we obtain the Mermin-Ho relation
\begin{equation}\label{eq:supptoprel}
    (\grad \times \bs w)_k = \frac{\epsilon_{ijk}}{2}\*s\cdot(\partial_i\*s\times\partial_j \*s),
\end{equation}
where the term on the right-hand side is the skyrmion density.

The Poisson brackets between $\{\rho,\*{s},\*{w}\}$ follow from the canonical Poisson brackets of Eq.~\eqref{eq:supppoissondvec}.
We find after some algebra
\begin{subequations}\label{supeq:newpoiss}
\begin{align}
\{s_i(\*{r}),s_j(\*r')\}&=\frac{1}{s}\epsilon_{ijk}s_k\delta(\*r - \*r'),\label{eq:supppoissonss}\\
\left\{w_i(\*r),w_j(\*r')\right\}&= \frac{1}{s}\epsilon_{ijk} \left(\grad \times \*w\right)_k \delta(\*r-\*r'), \\
\left\{w_i(\*r),\*s (\*r') \right\}&= \frac{1}{s} \partial_i \*s \,\delta(\*r-\*r'),\\
\left\{w_i(\*r),\rho(\*r^\prime)\right\}&=-q\partial_i \delta(\*r-\*r')/\hbar.\label{eq:supppoissonwrho}
\end{align}
\end{subequations}
Note that the Poisson bracket Eq.~\eqref{eq:supppoissonss} justifies the interpretation of $s\* s$ as the spin density with magnitude $s=\hbar d^2$. Similarly, the Poisson bracket Eq.~\eqref{eq:supppoissonwrho} allows the identification of $\rho=q d^2$ as the total charge density.  

\section{2. Dissipation}
In this section, we describe how dissipation can be included in the equations of motion while still maintaining certain topological constraints. For a noncanonical system of fields $\vb{X}=\{X_1,X_2,\ldots\}$, we can describe the dynamics using Hamilton's equation,
\begin{equation}
    \dot X_i(\vb{r},t) =\big\{X_i(\vb{r},t),\mc F\big\}= \int \mathrm{d}{V^\prime}J_{ij}(\vb{r},\vb{r}^\prime) {f}_j(\vb{r}',t). \label{eq:suppham}
\end{equation}
From here on, we assume the convention of Einstein summation over repeated indices. We also introduce the Poisson operator as $J_{ij}(\vb{r},\vb{r}^\prime)\equiv\big\{X_i(\vb{r},t),X_j(\vb{r}',t)\big\}$ and $f_j\equiv {\delta_{X_j} \mc F}$ as the force thermodynamically conjugate to the field ${X}_j$, where $\cal F[\vb{X}]$ the free energy functional. 

\subsection{(i) Casimir invariants}
Before we move on to introduce dissipation, we stress that the noncanonical Poisson structure can give rise to Casimir invariants $\mc Q$. 
Namely, if 
\begin{align}
\{\mc Q,{\mc {G}}\}=\int \mathrm{d}{V^\prime} {\int \mathrm{d}{V} \frac{\delta \mc Q}{ \delta {X_i}(\vb{r},t) } J_{ij}(\vb{r},\vb{r}^\prime)}\frac{\delta \mc G}{ \delta {X_i}(\vb{r}',t) } =0,
\end{align}
for any functional ${\mc G}$, the Casimir invariant $\mc Q$ is always conserved. The conserved charge has topological origins, as it only derives from the Poisson brackets rather than from symmetries of the free energy $\mc F$. Thus, it is distinct from the conserved Noether charge. Whenever such Casimir invariants exist, the Poisson operator $J_{ij}(\vb{r},\vb{r}^\prime)$ has a finite kernel, and, hence, no inverse. Importantly, this means that while the velocities $\dot{X}_i$ can always be expressed as a functional of the forces $f_i$ via Eq.~\eqref{eq:suppham}, the converse does not generally hold. This will become crucial when we include dissipation below. 

As an educational example, we consider a magnet with  $\*X=\{s_x, s_y,s_z\}$ describing the spin density $\vb{s}$. The noncanonical Poisson brackets fulfill $\{ s_i(\vb{r},t), s_j(\vb{r}^\prime,t)\}=\delta(\vb{r}-\vb{r}^\prime) \epsilon_{ijk} s_k(\vb{r},t)$. For the Casimir invariant $\mc Q= \vb{s}^2$, one can easily verify by using $\int \mathrm{d}V\delta_{s_i}\mc Q=s_i$ and $s_i J_{ij}=0$, that $\mc Q$ is conserved irregardless of the free energy $\mc F$. This explains why the equation of motion $\dot{\vb{s}}= \vb{s}\times \* f_{\vb{s}}$ cannot be inverted. 

\subsection{(ii) $\mc Q$-conserving dissipation}
For dissipation to conserve the Casimir invariants $\mc Q$, we propose the following equations of motion
\begin{equation}
    \dot X_i(\vb{r},t) = \int \mathrm{d}{V^\prime}J_{ij}(\vb{r},\vb{r}^\prime)\left[\frac{\delta \mc F [\vb{X}]}{\delta X_j(\vb{r}',t)}  +\frac{\delta \mc R[\dot{\vb{X}}]}{\delta \dot{X}_j(\vb{r}',t)} \right],\label{eq:suppdissvel}
\end{equation}
where the Rayleighian is constructed as a bilinear functional of the velocities
\begin{align}
    \mc R [\dot{\vb{X}}]= \int \mathrm{d}V\int \mathrm{d}V^\prime \frac{1}{2} \dot{X}_i (\vb{r},t) R_{ij}(\vb{r},\vb{r}^\prime) \dot{X}_j(\vb{r}^\prime,t).
\end{align}
Here, $R_{ij}(\vb{r},\vb{r}^\prime)=R_{ji}(\vb{r}^\prime,\vb{r})$ is  a local kernel that can depend on the fields $\vb{X}$. Note that, by construction, the dissipation in Eq.~\eqref{eq:suppdissvel} cannot violate the conservation of Casimir invariants, as it complies with the Poisson structure. 

Coming back to our example of the magnet, a Rayleighian $\mc R= \int \mathrm{d}V \alpha \dot{\vb{s}}^2/2$ will generate the equations of motion $\dot{\vb{s}}= -\vb{s}\times (\vb{f}_{\vb{s}} +\alpha \dot{\vb{s}})$, where it can be easily seen that, indeed, Gilbert damping cannot change the Casimir invariant $\mc Q=\vb{s}^2$. 

\subsection{(iii) $\mc Q$-nonconserving dissipation}
If we want to include dissipation that breaks the conservation law of Casimir invariants $\mc Q$, we have to use the more general version of the equations of motion given by~\cite{felderhofJCP11}
\begin{equation}
    \dot X_i(\vb{r},t) = \int \mathrm{d}{V^\prime}J_{ij}(\vb{r},\vb{r}^\prime) f_j(\vb{r}^\prime,t)  -\frac{\delta \mc R[\vb{f}]}{\delta f_i(\vb{r},t)}.\label{eq:suppdissforce}
\end{equation}
Here, the Rayleighian $\mc R[\vb{f}]$ is constructed as a functional of the thermodynamic forces $\vb{f}=\{\delta_{X_1}\mc F,\delta_{X_2}\mc F,\ldots\}$ given by
\begin{align}
    \mc R [{\vb{f}}]= \int \mathrm{d}V\int \mathrm{d}V^\prime \frac{1}{2} {f}_i (\vb{r},t) L_{ij}(\vb{r},\vb{r}^\prime) {f}_j(\vb{r}^\prime,t),
\end{align}
where $L_{ij}(\vb{r},\vb{r}^\prime)=L_{ji}(\vb{r}^\prime,\vb{r})$ are dissipation coefficents.
Only when the functional depends implicitly on the forces through velocities, $\mc R[\vb{f}]=\mc R[\dot{\vb{X}}(\vb{f})]$, can we use the chain rule 
\begin{align}
    -\frac{\delta \mc R[\dot{\vb{X}}(\vb{f})]}{\delta f_i(\vb{r},t)} =-\int \mathrm{d}V^\prime \frac{\delta \dot{X}_j(\vb{r}^\prime,t)}{\delta f_i(\vb{r},t)} \frac{\delta \mc R[\dot{\vb{X}}]}{\delta {\dot{X}_j}(\vb{r}^\prime,t)} \approx -\int \mathrm{d}V^\prime J_{ji}(\vb{r}^\prime,\vb{r}) \frac{\delta \mc R[\dot{\vb{X}}]}{\delta {\dot{X}_j}(\vb{r}^\prime,t)}= \int \mathrm{d}V^\prime J_{ij}(\vb{r},\vb{r}^\prime) \frac{\delta \mc R[\dot{\vb{X}}]}{\delta {\dot{X}_j}(\vb{r}^\prime,t)}
\end{align}
and retrieve the $\mc Q$-conserving dissipation of Eq.~\eqref{eq:suppdissvel}. In the second step, we have reinserted the equations of motion and, for simplicity, neglected higher orders in dissipation. In the third step, we use the antisymmetry of the Poisson operator, $J_{ji}(\vb{r}^\prime,\vb{r})=-J_{ij}(\vb{r},\vb{r}^\prime)$.

For our example of a magnet, if we instead use a Rayleighian constructed from the thermodynamic forces, $\mc R=\int \mathrm{d}V \alpha \, {\vb{f}_{\vb{s}}}^2/2$, the equations of motion would be $\dot{\vb{s}}= -\vb{s}\times \vb{f}_{\vb{s}} - \alpha {\vb{f}_{\vb{s}}}$. Here, $\* f_\*s$ can have components along $\*s$ so the length $\mc Q=\vb{s}^2$ is allowed to change. 
We retrieve Landau-Lifshitz damping (which conserves $\mc Q=\vb{s}^2$) by using  $\mc R=\int \mathrm{d}V \alpha (\vb{s}\times{\vb{f}_{\vb{s}}})^2/2$, which projects out the components of $\*f_\*s$ parallel to $\*s$. 

\subsection{(iv) Onsager symmetry}
When constructing the Rayleighian, we have to respect the second law of thermodynamics and microscopic reversibility. The former prescribes that $L_{ij}$ and $R_{ij}$ must be positive semi-definite matrices.
The latter is manifest as Onsager reciprocity~\cite{groot13}, which relates the coefficients via
\begin{align}
    L_{ij}(\vb{r},\vb{r}^\prime)=\epsilon_j \epsilon_i \mc T[L]_{ji}(\vb{r}^\prime,\vb{r}),
\end{align}
where $\mc T$ is the time-reversal operation. Thus, $\mc T[L]_{ji}$ transforms all (external and internal) fields that might be included in $L_{ji}$. In addition, $\mc T[X_i]=\varepsilon_i X_i$ defines the parity $\varepsilon_i=\pm 1$ under time reversal. Note that this relation ensures that the Rayleighian is even under time reversal:
\begin{align}
    \mc T[\mc R[\vb{f}]]=\int \mathrm{d}V\int \mathrm{d}V^\prime \frac{1}{2} {f}_j (\vb{r}^\prime,t) \epsilon_j \epsilon_i \mc T[L]_{ji}(\vb{r}^\prime,\vb{r}) {f}_i(\vb{r},t)=\mc R[\vb{f}],
\end{align}
where we used that $\mc T[f_i]=\varepsilon_i f_i$ since $f_i=\delta_{X_i}\mc F$ and $\mc T[\mc F]=\mc F$. 
Hence, a Rayleighian that is Onsager symmetric has to be even under time reversal. 
An analog statement holds for the Rayleighian as a functional of velocities, $\mc T[\mc R[\dot{\vb{X}}]]=\mc R[\dot{\vb{X}}]$, where $R_{ij}$ also fulfills the Onsager relations (as does the Poisson operator $J_{ij}$). 

\section{3. Equations of motion}

In the minimal model for a 3D triplet superconductor that includes the Zeeman coupling and allows for charge fluctuations, the free energy can be written in terms of the fields $\*X=\{\rho,\*s,\*w,\*A,\*E\}$, with the total charge density $\rho$, the spin orientation field ${\vb{s}}$, the superconducting winding density $\*w$, the magnetic vector potential $\*A$ and the electric field $\*E$. 
Once the Poisson brackets, the free energy $\mc F$, and the Rayleighian $\mc R$ are specified, we can use Eq.~\eqref{eq:suppdissvel} to derive the equations of motion. 

The Poisson brackets between $\{\rho,\*s,\*w\}$ are given by Eq.~\eqref{supeq:newpoiss}, which have to be supplemented by the canonical Poisson bracket between $\*A$ and $\*E$,
\begin{equation}\label{supeq:empoiss}
    \{E_i(\*r),A_j(\*r')\} = 4\pi c\,\delta_{ij}\delta(\*r - \*r'),
\end{equation}
to capture the electrodynamics. The remaining Poisson brackets vanish. 

The free energy can be phenomenologically written as 
\begin{align}\label{supeq:freeenergy}
    \mc F[\rho,\*s,\*w,\*A,\*E] &= \int \mathrm{d}V\left\{\frac{A}{2}\left(\grad\*s\right)^2 -\gamma\hbar\rho\* B\cdot \*s + \frac{1}{2\chi_w}\left(\* w - \frac{q}{\hbar c}\*A\right)^2 + \frac{(\rho - \rho_0)^2}{2\chi_c}  +\frac{\*{B}^2 + \*E^2}{8\pi}\right\}
\intertext{which has been written using the magnetic field}
    \*{B}&=\grad\times \*{A}.
\end{align}
Here, $A$ is the spin stiffness, $\chi_w$ is the winding compressibility, $\chi_c$ is the charge compressibility, and $\gamma<0$ is the gyromagnetic ratio. 
The connection between the free energy in Eq.~(\ref{supeq:freeenergy}) and the Ginzburg-Landau free energy for the order parameter $\vb{d}$ is elucidated by rewriting the kinetic term of the Ginzburg-Landau free energy in terms of the fields $\{\rho,\*{s},\*w\}$ via
\begin{align}\label{freeEnergy}
    \frac{A_d}{2}\left\vert \left(\grad -i\frac{q}{\hbar c}\vb{A}\right) \vb{d}\right\vert^2 =  \frac{A_d\rho}{4}  \left(\grad \vb{s}\right)^2 +\frac{A_d\rho}{2}\left(\* w - \frac{q}{\hbar c}\*A\right)^2 +\frac{A_d\rho}{2}\left(\grad \sqrt{\rho}\right)^2.
\end{align}
We find that this produces three terms which, respectively, are analogous to the spin exchange energy, the superconducting winding energy, and a term also referred to as ``quantum pressure"~\cite{bradleyPRA22} that is neglected in our analysis. Generically, we may expect that the prefactors of each of these terms are independent phenomenological parameters, rather than being related via $A_d$. This results in the spin exchange coefficient $A$ and the effective winding compressibility $\chi_w$.

Energy dissipation is governed by the Rayleigh functional $\mc R$, which is bilinear in the velocities $\dot{\*X}$
\begin{align}
    \mc R &= \int \t dV \left\{\frac{\alpha}{2}\hbar\rho(\partial_t\*s)^2 + \frac{\sigma}{2}\left[\frac{\hbar}{q}\partial_t\left(\*w - \frac{q\*A}{\hbar c}\right)\right]^2\right\},
\end{align}
where the first term corresponds to Gilbert damping with damping coefficient $\alpha$ and the second term describes Joule heating with conductivity $\sigma$. Now, we have all the ingredients to derive the equations of motion.

\subsection{(i) Charge continuity}
For the charge density $\rho$, there is one nonzero Poisson bracket
\begin{equation}
\begin{split}
    &\int \mathrm{d}{V'}\,\left\{\rho(\*{r}),w_j(\*{r}')\right\}\left[\frac{\delta  \mc F}{\delta w_j(\*{r}')}+\frac{\delta\mc R}{\delta \dot{X}_j(\*r')}\right]
    \\&=\int \mathrm{d}{V'}\,[\partial'_j\delta(\*r'-\*r)]\left[\frac{q}{\hbar\chi_w} \left(w_j(\*r') - \frac{q}{\hbar c}A_j(\*r')\right) +\frac{\hbar \sigma}{q} \left(\dot{w}_j(\*r') - \frac{q}{\hbar c}\dot{A}_j(\*r')\right)\right]\\
    &=-\grad\cdot\left[\frac{q}{\hbar\chi_w}\left(\*w - \frac{q\*A}{\hbar c}\right)+\frac{\hbar \sigma}{q}\left(\dot{\*w} - \frac{q\dot{\*A}}{\hbar c}\right)\right]\\
    &= -\grad\cdot\left(\*j +\sigma \bs{\mc{E}}\right),
\end{split}
\end{equation}
where in the last step we identified the supercurrent $\vb{j}=\frac{q}{\hbar\chi_w}\left(\*w - \frac{q\*A}{\hbar c}\right)$ and the normal current $\sigma \bs{\mc{E}}$ stemming from the total electromotive focre $\bs{\mc{E}}=\vb{E} + \hbar \dot{\*w}/q$.
The equation of motion becomes
\begin{align}\label{supeq:drhodt}
    \Aboxed{\partial_t \rho  = -\grad\cdot \bs{\mf{J}}},
\end{align}
where the total current is $\bs{\mf{J}}=\*j +\sigma \bs{\mc{E}}$.

\subsection{ (ii) Landau-Lifshitz-Gilbert equation}
For the spin orientation $\* s$, we have two nonzero Poisson brackets. The first Poisson bracket generates the term
\begin{equation}
\begin{split}
    &\int \mathrm{d}{V'}\,\left\{s_i(\*{r}),s_j(\*{r}')\right\}\left[\frac{\delta  \mc F}{\delta s_j(\*{r}')}+\frac{\delta\mc R}{\delta \dot{s}_j(\*r')}\right]
    \\&=\int \mathrm{d}{V'}\,\epsilon_{ijk}s_k(\*r)\delta(\*r - \*r')\left[\frac{1}{s(\*r')} \frac{\delta  \mc F}{\delta s_j(\*{r}')}+\alpha\dot{s}_j(\*r') \right]\\
    &=\epsilon_{ijk}s_j \left(\*H_\text{eff} - \alpha \dot{\*s}\right)_k,
\end{split}
\end{equation}
where we defined the effective field
\begin{align}
    \*H_\t{eff} &\equiv -\frac{1}{s} \frac{\delta \mc F}{\delta{\*s}} = \frac{A}{s}\grad^2\*s -\gamma\*B.
\end{align}
The second Poisson bracket generates the term 
\begin{equation}
\begin{split}
    &\int \mathrm{d}{V'}\,\left\{s_i(\*{r}),w_j(\*{r}')\right\}\left[\frac{\delta  \mc F}{\delta w_j(\*{r}')}+\frac{\delta\mc R}{\delta \dot{w}_j(\*r')}\right]
    \\&=-\int \mathrm{d}{V'}\frac{\hbar}{q s}[\partial_j's_i(\*r')]\delta(\*r - \*r'){\mf{J}}_j(\*r') \\
    &=- \frac{\bs{\mf{J}}}{\rho} \cdot \grad s_i,
\end{split}
\end{equation}
where we have identified the total current $\bs{\mf{J}}$ and used the relation between spin and charge density $\rho=q s/\hbar$. Hence, the Landau-Lifshitz-Gilbert equation is given by
\begin{align}
   \Aboxed{\partial_t\*s  &= \*s\times \*H_\t{eff} - \alpha\*s\times\partial_t\*s - \frac{\bs{\mf{J}}}{\rho} \cdot \grad \*s}.\label{eq:suppdsdt} 
\end{align}

\subsection{(iii) Winding continuity equation}
For the winding density $\* w$, we have three nonzero Poisson brackets. The first Poisson bracket generates the term
\begin{equation}
\begin{split}
    &\int \mathrm{d}{V'}\,\left\{w_i(\*{r}),w_j(\*{r}')\right\}\left[\frac{\delta  \mc F}{\delta w_j(\*{r}')}+\frac{\delta\mc R}{\delta \dot{w}_j(\*r')}\right]
    \\&=\int \mathrm{d}{V'}\,\frac{1}{\rho}\epsilon_{ijk} \left[\grad \times \*w(\*r)\right]_k \delta(\*r - \*r'){\mf{J}}_j(\*r')\\
    &=\frac{1}{\rho}\epsilon_{ijk}{\mf{J}}_j \left(\grad \times \*w\right)_k .
\end{split}
\end{equation}
The second Poisson bracket generates the term
\begin{equation}
\begin{split}
    &\int \mathrm{d}{V'}\,\left\{w_i(\*{r}),s_j(\*{r}')\right\}\left[\frac{\delta  \mc F}{\delta s_j(\*{r}')}+\frac{\delta\mc R}{\delta \dot{s}_j(\*r')}\right]
    \\&=-\int \mathrm{d}{V'}\,[\partial_i s_j(\*r)]\delta(\*r - \*r')\left[\*H_\text{eff}(\*r') - \alpha \dot{\*s}(\*r')\right]_j\\
    &=-(\partial_i s_j)\left(\*H_\text{eff} - \alpha \dot{\*s}\right)_j.
\end{split}
\end{equation}
Finally, the third Poisson bracket generates the term 
\begin{equation}
\begin{split}
    &\int \mathrm{d}{V'}\,\left\{w_i(\*{r}),\rho(\*{r}')\right\}\left[\frac{\delta  \mc F}{\delta \rho(\*{r}')}+\frac{\delta\mc R}{\delta \dot{\rho}(\*r')}\right]
    \\&=\int \mathrm{d}{V'}\,\frac{q}{\hbar}[\partial_i'\delta(\*r - \*r')]\mu(\*r')\\
    &=-\frac{q}{\hbar}\partial_i \mu,
\end{split}
\end{equation}
where we have defined the chemical potential
\begin{align}
    \mu = \frac{\delta  \mc F}{\delta \rho}=\frac{\rho-\rho_0}{\chi_c}-\gamma\hbar \*s\cdot\*B. 
\end{align}
Putting all together, the total equation becomes
\begin{align}
    \partial_t\* w=-\frac{q}{\hbar}\grad \mu +\frac{\bs{\mf{J}}}{\rho}\times\left(\grad\times\*w\right) -  \left\{\*{s} \times\left[\left(\*H_\text{eff} - \alpha \dot{\*s} \right) \times \*s\right] \right\}\cdot \grad \*s,
\end{align}
where for the last term we have used that $\grad \*s$ is orthogonal to $\vb{s}$, allowing us to introduce the projection $\*{s} \times \left( \ldots \times \*s\right)$. Reinserting the Landau-Lifshitz-Gilbert equation from Eq.~\eqref{eq:suppdsdt}, we get
\begin{align}
    \partial_t\* w&=-\frac{q}{\hbar}\grad \mu +\frac{\bs{\mf{J}}}{\rho}\times\left(\grad\times\*w\right) + \left[\*{s} \times \left(\partial_t \*s +\frac{\bs{\mf{J}}}{\rho}\cdot\grad \*s\right)\right] \cdot \grad \*s.
\end{align}
Making use of the topological equation from Eq.~\eqref{eq:supptoprel}, we obtain 
\begin{align}
 \Aboxed{\partial_t\* w=-\frac{q}{\hbar}\grad \mu  +\*s \cdot \left(\partial_t \*s \times \grad \*s \right) }. \label{eq:suppwindingeq}
\end{align}
Note that the equation of motion can be rewritten as a spin-triplet version of the first Josephson relation
\begin{align}
    \partial_t \*{j} = \frac{q^2}{\hbar^2\chi_w}\bs{\mc{E}},
\end{align}
where the electromotive force is $\bs{\mc{E}}=\vb{E}-\grad \mu+ \hbar\*s \cdot \left(\partial_t \*s \times \grad \*s \right)/q $. Similarly, the spin-triplet version of the second Josephson relation becomes with Eq.~\eqref{eq:supptoprel}
\begin{align}
    \grad\times\*{j}=-\frac{q^2}{\hbar^2\chi_w c}\bs{\mc{B}},
\end{align}
where the total effective magnetic field is given by $\bs{\mc{B}}=\*B + \hbar c {\epsilon_{ijk}} \*{s} \cdot \left(\partial_j \*s \times \partial_k \*s \right)/2q$.
\subsection{(iv) Ampère's law}
For the electric field $\*E$, there is only one nonzero Poisson bracket
\begin{equation}
\begin{split}
    &\int \mathrm{d}{V'}\,\left\{E_i(\*{r}),A_j(\*{r}')\right\}\left[\frac{\delta  \mc F}{\delta A_j(\*{r}')}+\frac{\delta\mc R}{\delta \dot{A}_j(\*r')}\right]
    \\&=-\int \mathrm{d}{V'}\,4\pi \delta_{ij}  \delta(\*r - \*r')\left\{{\mf{J}}_j(\*r')+ c  \epsilon_{jkl} \partial_k' \left[\gamma \hbar \rho(\*r') s_l(\*r')\right]-\frac{c}{4\pi} \epsilon_{jkl}\partial_k' \left[\epsilon_{l mn}\partial_m' A_n(\*r')\right]\right\}\\
    &=\left( -4\pi\bs{\mf{J}}-4\pi c\grad\times\*M+c\grad\times \*B\right)_i,
\end{split}
\end{equation}
where we defined the magnetization $\*M=\gamma \hbar \rho \*s$. Hence, the equation becomes
\begin{align}
    \Aboxed{\partial_t\*{E}=-4\pi\bs{\mf{J}} +c\grad\times \left(\*B -4\pi\*M\right)}.
\end{align}
This confirms that $\bs{\mf{J}}$ is the total charge current in the system. Furthermore, the equation for $\*{A}$ just reproduces the Weyl gauge, $\partial_t \*{A}= -c \*{E}$. While $\grad\cdot\*B=0$ is automatically fulfilled, Gauss's law $\grad\cdot\*E=4\pi \rho$ has to be imposed as an additional constraint. 

\section{4. Topological hydrodynamics}
\subsection{(i) Winding charge in 1D}
For a 1D wire, the topological continuity equation of winding is given by the 1D version of Eq.~\eqref{eq:suppwindingeq},
\begin{align}\label{eq:1dcont}
 \partial_t w=-\frac{q}{\hbar}\partial_\ell \mu + j_\text{sk},
 \end{align}
 where  we define the skyrmion flux $j_\t{sk}=\*s \cdot \left(\partial_t \*s \times\partial_\ell \*s \right)$ transverse to the wire. When magnetic dynamics are frozen, $j_\t{sk} =0$, the topological hydrodynamics of superconducting winding stemming from the 1D bulk homotopy $\pi_1(S^1)=\mb Z$ is restored. On topological grounds, it is therefore known that winding satisfies the continuity equation 
 \begin{align}
      \partial_tw = -\partial_\ell j_w,
 \end{align}
where the winding flux is given by $j_w=-\left(\partial_t\phi+\*e_1\cdot\partial_t\*e_2\right)$, see Eq.~\eqref{eq:winding}. Inspecting the continuity equation in Eq.~(\ref{eq:1dcont}), the winding flux can be identified with the gradient of the chemical potential via $j_w=q\partial_\ell \mu/\hbar$. 
On the other hand, if $ j_\t{sk}\neq 0$, the magnetic dynamics can induce nonsingular phase slips in the winding, which is the key mechanism used in the SQUID device presented in the Letter. 
\subsection{(ii) Skyrmion charge in 2D}
For a 2D system, the topological continuity of magnetic skyrmions follows from rewriting Eq.~\eqref{eq:suppwindingeq} into  
\begin{align}
 \partial_t\* w=-\frac{q}{\hbar}\grad \mu  -\hat{\* z}\times \* j_\text{sk},
\end{align}
where the skyrmion flux is given by $\* j_\t{sk}=\*s \cdot \left[\partial_t \*s \times \left(\hat{\* z}\times\grad\right) \*s \right]$. Calculating the rotation of the equation and using that $\hat{\* z}\cdot \grad\times \*w=\rho_\t{sk}$ is the skyrmion density, we immediately obtain the skyrmion continuity equation
\begin{align}
 \partial_t \rho_\t{sk}=  -\grad\cdot \* j_\t{sk}.
\end{align}
In fact, the skyrmion charge $\mc Q=\int \mathrm{d}r^2\rho_\t{sk}$ is a Casimir invariant regardless of the form of $\cal F[\* X]$ and $\cal R[\dot{\* X}]$.
Note, the skyrmion continuity equation is also fulfilled in conventional Heisenberg magnets, which have an $S^2$ order parameter space. In that context, the continuity equation stems from the bulk homotopy $\pi_2(S^2) = \mb Z$, rather than from a bulk-edge correspondence described by $\hat{\* z}\cdot \grad\times \*w=\rho_\t{sk}$~\cite{zouPRB19}. Therefore, those skyrmions are local objects that may fluctuate in and out of existence. In contrast, skyrmions in spin-triplet superconductors are inherently nonlocal objects that cannot be removed by local surgery and, thus, exhibit long-range interactions~\cite{knigavkoPRB99}.

\subsection{(iii) Hopf charge in 3D}
In 3D, we can derive the topological continuity equation for the Hopf charge~\cite{liuPRL22} density $\rho_\t{H}=\*w\cdot (\grad\times\*w)$. After some algebra, we find from Eq.~\eqref{eq:suppwindingeq}
\begin{align}
     \partial_t\rho_\t{H}=-\grad\cdot\left[\*w\times\frac{q}{\hbar}\left(\grad \mu+\*e\right)\right]  +\frac{2\* e\cdot \*b}{c},
\end{align}
where we defined the emergent electric field $e_i=q\*s \cdot \left(\partial_t \*s \times \partial_i \*s \right)/\hbar$ and the emergent magnetic field $b_i=\hbar c{\epsilon_{ijk}}\*s\cdot(\partial_i\*s\times\partial_j \*s)/2q$. 
We show that the source term $\*e\cdot \*b$ is identically zero by rewriting it as
\begin{equation}
\begin{split}
    \frac{2\* e\cdot \*b}{c}&=\frac{1}{4} \epsilon^{\mu\nu\sigma\rho}\left[\*s\cdot(\partial_\mu\*s\times\partial_\nu \*s)\right]\left[\*s\cdot(\partial_\rho\*s\times\partial_\sigma \*s)\right]\\
    &=\frac{1}{4}\epsilon^{\mu\nu\sigma\rho}(\partial_\mu\*s\times\partial_\nu\*s)\cdot(\partial_\rho\*s\times\partial_\sigma\*s)\\
    &=\frac{1}{2}\epsilon^{\mu\nu\sigma\rho}(\partial_\mu\*s\cdot\partial_\rho\*s)(\partial_\nu\*s\cdot\partial_\sigma\*s)=0
\end{split}
\end{equation}
where the Greek letters run over space and time indices, $\mu\in\{0,1,2,3\}$. In the second line, we have used the fact that $\partial_{\mu}\*s\times\partial_{\nu}\*s\propto\*s$ to rewrite the scalar product as a dot product. Finally, in the third line, we use that the contraction of the symmetric term $\partial_\mu\*s\cdot\partial_\rho\*s$ with the antisymmetric Levi-Civita symbol $\epsilon^{\mu\nu\sigma\rho}$ yields zero. Thus, we obtain the conservation law of Hopf charge in 3D
\begin{align}
    \partial_t\rho_\t{H}=-\grad\cdot{\*j}_\t{H},
\end{align}
where ${\*j}_\t{H}=\*w\times\frac{q}{\hbar}\left(\grad \mu+\*e\right)$.
For a fluid, this conservation law corresponds to hydrodynamic helicity by identifying the winding $\*w$ with the velocity field. There, it is a measure of the knottedness of vortex lines in the flow. Similarly, here, it is a measure of the knottedness of skyrmion flux tubes like Hopfions. We remark that the term $\*e\cdot \*b$ can be nonzero if the magnitude of $\*s$ is allowed to vary. Therefore, the presence of magnetic hedgehog dynamics would spoil the conservation law, since at the core of the hedgehog, the magnitude of $\*s$ smoothly goes to zero.

\section{5. Stability analysis}
To analyze the stability of our quasistatic ansatz for the 1D ring used in the Letter, we linearize the equations of motion
\begin{subequations}
\begin{align}
  \partial_t \rho  &= -\partial_\ell  (j +\sigma {\mc{E}}),\\
     \partial_t\*s  &= \*s\times \*H_\t{eff} - \alpha\*s\times\partial_t\*s - \frac{j +\sigma {\mc{E}}}{\rho} \partial_\ell \*s ,\\
    \partial_t {j} &= \frac{q^2}{\hbar^2\chi_w}{\mc{E}},
\end{align}
\end{subequations}
by using
\begin{equation}
\begin{split}
    \rho(\ell,t)&\rightarrow \rho+\delta \rho(\ell,t),\\
     \*s(\ell,t) &\rightarrow \*s(\ell) + \delta \*s(\ell,t),\\
    j(\ell,t)&\rightarrow j +\delta j(\ell,t).
\end{split}
\end{equation}
Our zeroth-order ansatz is given by a constant current $j$, a constant charge density $\rho$, and the spin texture $\* s(\ell)$. 
Linearizing the equations of motion, we obtain the eigenvalue equation
\begin{align}
    \omega \,\delta \*{X}(k) = \Omega(k)  \,\delta \*X(k),
\end{align}
for the eigenmodes with frequency $\omega$. Here, the matrix $\Omega$ is a function of the wave number $k$, which, for the ring, only takes discrete values $k=m2\pi/L$ with $m\in\mathbb{Z}$. In the following, we discuss separately fluctuations for the regimes with collinear and noncollinear spin texture. 

\subsection{Collinear spin texture}
For the collinear spin texture, we assume in zeroth order $\*s=\*e_z$ with fluctuations $\delta \*s=\delta s_x\*e_x+\delta s_y\*e_y$.
Then, the matrix becomes 
\begin{align}
\Omega=
\left(
\begin{array}{cccc}
 -i\frac{\alpha  A k^2}{s }+\frac{ j k}{\rho} & i\frac{A k^2 }{s}+\frac{\alpha  j k}{\rho} & 0 & 0 \\
 -i\frac{A k^2}{s}-\frac{\alpha  j k}{\rho} & -i\frac{\alpha  A k^2}{s}+\frac{j k}{\rho} & 0 & 0 \\
 0 & 0 & 0 & \frac{ k q^2}{\chi_c \chi_w \hbar ^2} \\
 0 & 0 &  k & -i\frac{k^2 \sigma }{\chi_c} \\
\end{array}
\right)
\end{align}
in the basis $(\delta s_x, \delta s_y, \delta j,\delta \rho)$. 
We obtain for the magnons the quadratic dispersion
\begin{align}
    \omega(k)=(Ak^2/s+ j k/\rho)(1-i\alpha),
\end{align}
which is shifted by the supercurrent $j$ due to the spin transfer torque. 
For the charge and current degrees of freedom, we obtain the linear dispersion $\omega_q= k q/\hbar\sqrt{\chi_c \chi_w}-ik^2 \sigma/\chi_c+{\cal O}(\sigma^2)$. Note the caveat that a full treatment of fluctuations in $\*E$ and $\*B$ would gap these modes (in accordance with the Higgs mechanism) to $\omega_q\rightarrow \sqrt{\omega_q^2+\omega_p^2}$, where the plasma frequency is given by $\omega_p^2=4\pi q^2/\hbar^2\chi_w$. Hence, in the following, we are only concerned with the low-energy magnon modes. 

\begin{figure}
    \centering
    \includegraphics[width=0.95\linewidth]{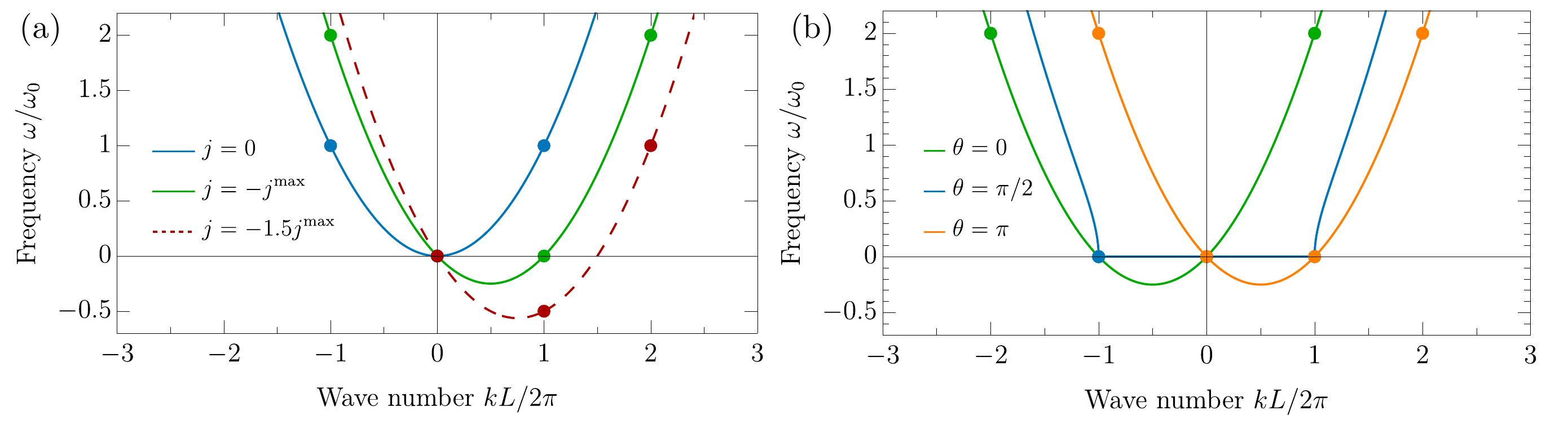}
    \caption{Magnon excitation spectrum for a (a) collinear spin texture and (b) noncollinear spin texture. In (a), once the current is lower than $j=-j^\text{max}$, the mode at $k=2\pi/L$ becomes unstable (orange dot). In (b), the quadratic dispersion is shifted from right (green) to left (orange)  as the phase slip is carried out. At $\theta=\pi/2$, the range $-2\pi/L<k<2\pi/L$ has a finite and positive imaginary part (not shown). However, at the allowed values $k=0$ and $k=\pm 2\pi/L$, the imaginary part is always zero. We defined $\omega_0=A(2\pi)^2/L^2s$.}
    \label{fig:spectrum}
\end{figure}

For stability, we have to ensure that $\t{Im}\,\omega(k)<0$. To satisfy this stability condition, we find that the current can only be increased up to
\begin{align}
    j=-A \frac{q}{\hbar} \frac{2\pi}{L}=-\kappa j_0=-j^\text{max}
\end{align}
before the dispersion dips too low, see Fig.~\ref{fig:spectrum}(a), and the lowest-energy mode at $k=2\pi/L$ becomes unstable. In the second step, we have used the definitions $j_0=2\pi q/\hbar \chi_w L$ and $\kappa=A\chi_w$ from the Letter. Once the instability is reached, this initiates the nonsingular phase slip. Note, introducing easy-axis anisotropy would modify the value of the maximal current. This is discussed in a later section of the Supplemental Material. 

\subsection{Noncollinear spin texture}
During the nonsingular phase slip, the spin texture traces out a skyrmion and, therefore, possesses a noncollinear spin texture. In zeroth order, we use our ansatz of a 
conical spin texture
\begin{align}\label{eq:suppspintext}
    \*s(\ell)=\begin{pmatrix}
\sin\theta \cos\varphi(\ell) \\
\sin\theta \sin\varphi(\ell) \\
\cos\theta
\end{pmatrix}, \quad   \*e_\theta(\ell)=\begin{pmatrix}
\cos\theta \cos\varphi(\ell) \\
\cos\theta \sin\varphi(\ell) \\
-\sin\theta
\end{pmatrix},\quad   \*e_\varphi(\ell)=\begin{pmatrix}
-\sin\varphi(\ell) \\
\cos\varphi(\ell) \\
0
\end{pmatrix},
\end{align}
where $\varphi(\ell)=p \ell$ and $p=2\pi l/L$.
The linearized fluctuations are described by
\begin{align}
\delta\*s(\ell,t)=\delta\theta(\ell,t)\,\*e_\theta(\ell) + \delta\varphi(\ell,t) \sin\theta\,\*e_\varphi(\ell).
\end{align}
In this case, the Landau-Lifshitz-Gilbert equation gives rise to a zeroth-order condition that the spin torque must be zero:
\begin{align}
    j=-\frac{q}{\hbar}  p \cos(\theta). 
\end{align}
Using this relation, we obtain for the fluctuations
\begin{align}
\Omega=
    \left(
\begin{array}{cccc}
 \frac{A q \left[k p \cos \theta-i \alpha  \left(k^2-p^2 \sin ^2\theta\right)\right]}{\rho \hbar } & \frac{A k q \sin \theta (\alpha  p \cos \theta+i k)}{\rho \hbar } & -\frac{i \alpha  p \sin \theta}{\rho} & 0 \\
 \frac{A \left[-i k^2 q \rho \csc \theta+p^2 \sin \theta (-k p \sigma  \hbar  \cos \theta+i q \rho)-\alpha  k p q \rho \cot \theta\right]}{\rho^2 \hbar } & \frac{A k \left[p q \rho \cos \theta-i k \left(p^2 \sigma  \hbar  \sin ^2\theta+\alpha  q \rho\right)\right]}{\rho^2 \hbar } & -\frac{i p}{\rho} & -\frac{k p \sigma }{\rho \chi_c} \\
 \frac{A p q^2 \sin \theta \left[k p \cos \theta-i \alpha  \left(k^2-p^2 \sin ^2\theta\right)\right]}{\rho \chi_w \hbar ^2} & \frac{A k p q^2 \sin ^2\theta (\alpha  p \cos \theta+i k)}{\rho \chi_w \hbar ^2} & -\frac{i \alpha  p^2 q \sin ^2\theta}{\rho \chi_w \hbar } & \frac{k q^2}{\chi_c \chi_w \hbar ^2} \\
 -\frac{i A k^2 p^2 \sigma  \sin \theta \cos \theta}{\rho} & \frac{A k^3 p \sigma  \sin ^2\theta}{\rho} & k & -\frac{i k^2 \sigma }{\chi_c} \\
\end{array}
\right)
\end{align}
in the basis $(\delta \theta, \delta \phi, \delta j,\delta \rho)$. 

Here, we find in the limit of zero dissipation, $\sigma=\alpha=0$, and vanishing charge compressibility, $\chi_c\rightarrow 0$, the dispersion
\begin{align}
    \omega(k)=\frac{2\pi A q}{\hbar\rho L}\left(k \cos\theta+\sqrt{\frac{L^2k^4}{4\pi^2}-k^2\sin^2\theta} \right).
\end{align}
As the nonsingular phase slip is carried out from $\theta=0$ to $\theta=\pi$, the magnon dispersion shifts from left to right, see Fig.~\ref{fig:spectrum}(b).

Including both the dissipation and compressibility again, it is insightful to study the modes with zero frequency $\omega=0$. First, we inspect the $k=0$ modes. We find that the only mode with a nonzero imaginary part is 
\begin{align}
\delta \* X(k=0)\propto
    \begin{pmatrix}
\alpha \sin\theta \\
1\\
\frac{\alpha  p q \sin ^2\theta }{\hbar\chi_w  }\\
0
\end{pmatrix}, \quad \omega(k=0)=-i\frac{ \alpha  p^2 q (1-\kappa) \sin ^2\theta}{s \chi_w }.
\end{align}
The imaginary frequency indicates that stability is ensured for $\kappa<1$. If $\kappa>1$, this mode corresponds to a runaway nonsingular phase slip that cannot be stabilized at any time. Thus, the noncollinear spin texture becomes unstable for $\kappa>1$. Nevertheless, the system can still work as a SQUID with abrupt nonsingular phase slips, see next section. 

\begin{figure}
    \centering
    \includegraphics[width=0.2\linewidth]{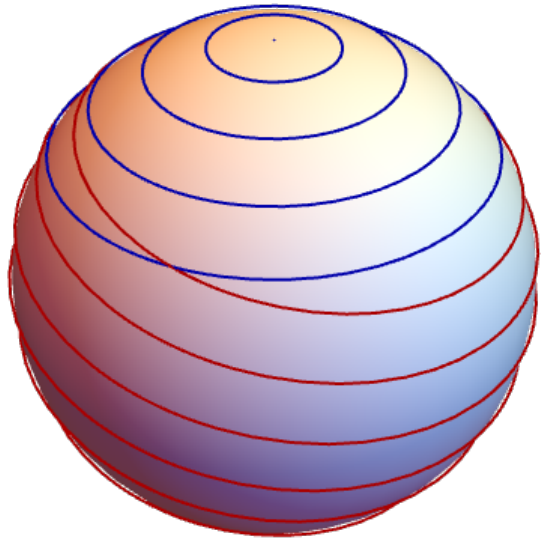}
    \caption{Mapping of spin texture of ring onto circle on the sphere, where each point on the sphere represents the direction of $\vb{s}$. For the nonsingular phase slip discussed in the Letter, the ring sweeps out the sphere symmetrically around the $z$-axis (blue circles). However, the system is susceptible to fluctuations that can change this symmetry axis (blue to red circles). This is confirmed in the magnon analysis, where we find a zero mode at $k=2\pi/L$ responsible for these fluctuations.  }
    \label{fig:placeholder}
\end{figure}

Another $\omega=0$ mode exists for $k=2\pi/L$ if the spin winding is $\abs{l}=1$. For larger spin winding $\abs{l}\geq2$, the $k=2\pi/L$ mode has a positive imaginary part, making the system unstable. Hence, only $l=0,\pm 1$ renders a stable ground state. Using $l=\pm 1$, we find the zero mode
\begin{align}
\delta \* X(k=\pm 2\pi/L)\propto
    \begin{pmatrix}
 \sin\theta \\
\pm i l \cos\theta\\
0\\
0
\end{pmatrix}, \quad \omega(k=\pm 2\pi/L)=0.
\end{align}
These modes maintain both the skyrmion charge and the exchange energy of the system. They merely change the symmetry axis of the conical spin texture, which for our ansatz is the $z$ axis. By mapping the spin texture of the ring onto circles on the sphere, where each point indicates its spin orientation, see  Fig.~\ref{fig:placeholder}, these fluctuations rigidly slide the circles around without deformation (from blue to red).   

\section{6. SQUID for $\kappa>1$ and with easy-axis anisotropy}
In the Letter, the main focus was on the $\kappa<1$ regime, in which the kinetic energy of the supercurrent dominates over the spin exchange energy. Furthermore, we assumed (for simplicity) that there were no anisotropies present, e.g., easy-axis anisotropy.  In this regime, the free energy analysis suggested that the nonsingular phase slip progresses quasistatically as the magnetic flux is varied and the spin texture traces out a skyrmion. 

However, in the $\kappa>1$ regime, the quasistatic ansatz is no longer maintained. Nevertheless, as the magnetic flux through the superconducting loop is increased, the supercurrent kinetic energy builds up as a diamagnetic response, which, through the bulk-edge correspondence, relaxes by undergoing a nonsingular phase slip and passing through a skyrmion. These nonsingular phase slips will be a ``runaway" process, since in the $\kappa>1$ regime the only possible stable fixed points for the magnetic texture are those in which the spin texture is collinear. Adding easy-axis anisotropy would also play a similar role in facilitating runaway nonsingular phase slip, since it would energetically prefer a collinear spin texture. Similarly to the Letter, we can perform a free energy analysis to characterize the magnetic and current response in the $\kappa>1$ regime with the addition of easy axis anisotropy to the free energy. 

\begin{figure}
    \centering
    \includegraphics[width=0.8\linewidth]{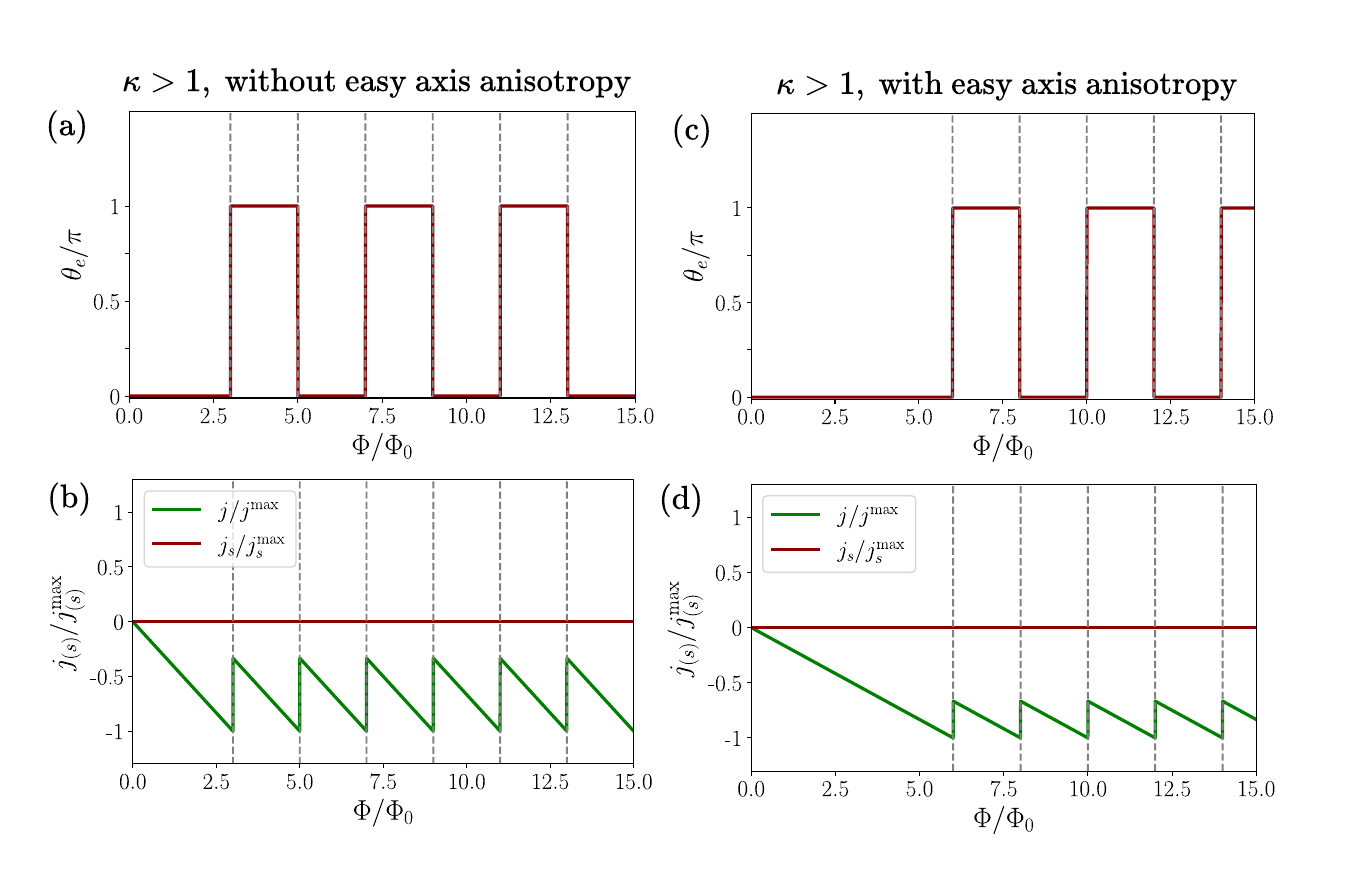}
    \caption{Magnetic and current response of the spin triplet SQUID in the $\kappa>1$ regime. (a) and (b) are plotted for $\kappa = 3$ and $K/L_kj_0^2=0$, while (c) and (d) are plotted for $\kappa = 3$ and $K/L_kj_0^2=3$. }
    \label{fig:kg1}
\end{figure}

To this end, the free energy can be given by
\begin{equation}
 \mc F_{nl}(\Phi,\theta) = \frac{1}{2}L_kj(\Phi,\theta)^2 + \frac{1}{2}Jl^2\sin^2\theta - \frac{K}{2} \cos^2\theta,
\end{equation}
where $K>0$ is the easy axis anistropy and $j(\Phi,\theta) = j_0[2n + l(1-\cos\theta) - \Phi/\Phi_0]$. Let us take the identical initial state as that presented in the Letter, and initialize the system so that the spin texture is collinear with $\*s = \*e_z$ and the magnetic flux is zero. As the magnetic flux is increased, see Figs.~\ref{fig:kg1}{\color{red}(a)} and ~\ref{fig:kg1}{\color{red}(c)}, the pole at $\theta = 0$ is stable up until the critical value of 
\begin{equation}
     \Phi_c = \left(\kappa + \frac{K}{L_kj_0^2}\right)\Phi_0,
\end{equation}
where $\Phi_0$ is the magnetic flux quantum. At this point, the current reaches its maximal magnitude given by 
\begin{equation}
    -j^\t{max} = -j_0\Phi_c/\Phi_0 = -j_0\left(\kappa + \frac{K}{L_kj_0^2}\right),
\end{equation}
see Figs.~\ref{fig:kg1}{\color{red}(b)} and ~\ref{fig:kg1}{\color{red}(d)}. Here, we find that without the easy axis anisotropy, the maximal current and the critical flux written in terms of $\kappa$ are the same as those of the Letter. Including the easy axis anisotropy stabilizes the collinear spin texture, thus requiring a higher magnetic flux, and hence a higher maximal current, before we reach an instability.

Once the pole at $\theta=0$ destabilizes, the runaway nonsingular phase slip occurs, during which the current relaxes by a fixed amount 
\begin{equation}
    \Delta j = 2j_0
\end{equation}
and the magnetic texture passes through a skyrmion, switching to $\theta= \pi$. This is reflected in Fig.~\ref{fig:kg1} via the sawtooth current response and the square wave magnetic response. As previously stated, we also see in Figs. \ref{fig:kg1}{\color{red}(c)} and \ref{fig:kg1}{\color{red}(d)} that the inclusion of easy axis anisotropy shifts the critical flux to a larger value. 

Note, in the analysis of the $\kappa<1$ regime presented in the Letter, the current is symmetric about $j=0$ and the magnitude $\Delta j = 2\kappa j_0$ is smaller. As $\kappa\rightarrow 1$, the change in current saturates to $\Delta j \rightarrow 2j_0$. This occurs because the energy relaxation corresponding to passing through a single skyrmion has a maximal amount. After the current is relaxed, further increasing the magnetic flux raises the current magnitude back to $j^\t{max}$, whereupon the process repeats with periodicity $2\Phi_0$, which is the same periodicity as in the Letter.

Finally, it is important to note that the average current being nonzero in the $\kappa>1$ regime is an indicator of spin-triplet superconductivity. While the sawtooth behavior of the current is reminiscent of that of the U(1) superconductor in the Little Parks effect~\cite{tinkham95}, there, the average current is zero. Here, the average current is nonzero because the supercurrent is topologically linked to the spin texture. The supercurrent can only relax once it becomes energetically preferred to distort the spin texture to pass through a skyrmion. Furthermore, here the $2\Phi_0$ periodicity is still maintained even when there is anisotropy, whereas for singular phase slips in U(1) superconductors the periodicity is $\Phi_0$.

\bibliography{supp.bib}